\documentclass[
  a4paper,
  superscriptaddress,
  twocolumn,
  prx,
  preprintnumbers,
  floatfix,
  secnumarabic,
  amssymb,
  nofootinbib
]
{revtex4-2}
\usepackage{lineno}
\usepackage{graphicx}  
\usepackage{dcolumn}   
\usepackage{bm}        
\usepackage{amssymb}   
\usepackage{amsmath,stmaryrd}
\usepackage{blkarray, multirow, graphicx, diagbox, color, xcolor, colortbl}
\usepackage{bbm, bbold}
\usepackage{glossaries}
\usepackage{hyphenat}
\usepackage{ifthen}
\usepackage{xkeyval}
\usepackage{moreverb}
\usepackage{rotating}
\usepackage{wrapfig}
\usepackage{slashbox}
\usepackage{xspace}
\usepackage{nicefrac}
\usepackage[]{units}
\usepackage{physics}
\usepackage{booktabs}
\usepackage{braket}
\usepackage[inline]{enumitem}
\usepackage{tabto}
\usepackage{listings}
\usepackage{xstring}
\usepackage{lipsum}

\def\ReplaceStr#1{%
	\IfSubStr{#1}{p}{%
		\StrSubstitute{#1}{p}{.}}{#1}}

\usepackage[caption=false]{subfig}
\usepackage[customcolors]{hf-tikz} 
\usepackage{tikz}
\usepackage{calc}
\usetikzlibrary{svg.path}
\usepackage{pgffor}
\usepackage{pgfplots}
\pgfplotsset{compat=1.13}
\usepackage{pgfplotstable}
\usepgfplotslibrary{groupplots}
\usepgfplotslibrary{fillbetween}

\tikzstyle{n} = [draw,shape=ellipse,minimum size=1.5em,inner sep=0pt,fill=white!20, minimum width=2.5em]
\tikzstyle{Init} = [n,color=green,fill=green!20,text=black]
\tikzstyle{Fin} = [n,color=red,fill=red!20,text=black]
\tikzstyle{Ghost} = [minimum size=1.5em,inner sep=0pt,color=white,text=black]
\tikzstyle{Multiple} = [draw,shape=rect,minimum size=2em,inner sep=0pt]

\tikzstyle{ghostA} = [text=red!70,thick, minimum size=2*(5pt-\pgflinewidth), inner sep=0pt, outer sep=0pt]
\tikzstyle{ghostB} = [text=blue!70,thick, minimum size=2*(5pt-\pgflinewidth), inner sep=0pt, outer sep=0pt]
\tikzstyle{siteA} = [regular polygon, regular polygon sides=3, shape border rotate= 30, draw=red!50,fill=red!20,thick,inner sep=0pt,minimum width=1.5em,font=\footnotesize]
\tikzstyle{siteB} = [regular polygon, regular polygon sides=3, shape border rotate= -30, draw=green!50,fill=green!20,thick,inner sep=0pt,minimum width=1.5em,font=\footnotesize]
\tikzstyle{op} = [regular polygon, regular polygon sides=4, draw=orange!50, fill=orange!20, thick, inner sep=0.2pt, minimum width=1.25em, minimum height=1.5em,font=\footnotesize]
\tikzstyle{opghost} = [regular polygon, regular polygon sides=4, thick, inner sep=0.2pt, minimum width=1.25em, minimum height=1.5em,font=\footnotesize]
\tikzstyle{site} = [circle,draw=blue!50,fill=blue!20,thick,inner sep=0.2pt,minimum width=1.25em,font=\footnotesize]
\tikzstyle{hiddensite} = [circle,draw=white!50,fill=white!20,thick,inner sep=0.2pt,minimum width=1.25em,font=\footnotesize]
\tikzstyle{nosite} = [circle,draw=white,fill=white,thick,inner sep=0.1pt,minimum width=1.5em]
\tikzstyle{ghost} = [font=\footnotesize]
\tikzstyle{intersite} = [regular polygon, regular polygon sides=4, shape border rotate= 45, draw=black!50,fill=black!20,thick,inner sep=0pt,minimum width=1.5em]
\tikzstyle{ld} = [inner sep=1pt, font=\small]
\tikzstyle{unsite} = [circle, outer sep=0pt,inner sep=0.2pt,minimum width=1.25em]

\definecolor{colorA}{rgb} {0.58,0,0.8275}
\definecolor{colorB}{rgb} {0.11,0.663,0.51}
\definecolor{colorC}{rgb} {0.3373,0.7059,0.9137}
\definecolor{colorD}{rgb} {0.902,0.6235,0}
\definecolor{colorE}{rgb} {0.9451,0.902,0.3255}
\definecolor{colorF}{rgb} {0.3373,0.3255,0.902}
\definecolor{colorG}{rgb} {0.9451,0.3255,0.3373}
\definecolor{cbColorA}{HTML} {810F7C}
\definecolor{cbColorB}{HTML} {006D2C}
\definecolor{cbColorC}{HTML} {7BCCC4}
\definecolor{cbColorD}{HTML} {C51B8A}
\definecolor{cbColorE}{HTML} {0868AC}

\usetikzlibrary
{
	calc,
	decorations,
	plotmarks,
	patterns,
	positioning,
	petri,
	arrows,
	decorations.markings,
	backgrounds,
	fit,
	graphs,
	shapes.geometric,
	decorations.pathmorphing,
	shapes.misc,
	shapes,
	tikzmark,
	pgfplots.colorbrewer,
	fpu,
}

\usepackage{ocgx}

\usetikzlibrary{ocgx}

\pgfplotsset{
        cycle from colormap manual style/.style={
            x=3cm,y=10pt,ytick=\empty,
            stack plots=y,
            every axis plot/.style={line width=2pt},
        },
}

\tikzset{>=stealth}

\tikzset{->-/.style={decoration={
			markings,
			mark=at position .5 with {\arrow{>}}},postaction={decorate}}}

\tikzset{-<-/.style={decoration={
			markings,
			mark=at position .5 with {\arrow{<}}},postaction={decorate}}}

\tikzstyle{orientedsnake} = [
decorate, 
decoration={snake},
->
]  
\tikzstyle{orientedshortarrow} = [
decoration={markings,
	mark=at position .33 with {\arrow{>}}},
postaction={decorate}
]  
\tikzstyle{orientedlongarrow} = [
decoration={markings,
	mark=at position .67 with {\arrow{>}}},
postaction={decorate}
]
\tikzset{dbl/.style={double,
		double equal sign distance,
		-implies,
		shorten >=10pt,
		shorten <=10pt}}
\tikzset{
	between/.style args={#1 and #2}{
		at = ($(#1)!0.5!(#2)$)
	}
}

\tikzstyle{process} = [rectangle, minimum width=3cm, minimum height=1cm, text centered, text width=5cm, draw=black]
\tikzstyle{io} = [trapezium, trapezium left angle=70, trapezium right angle=110, minimum width=3cm, minimum height=1cm, text centered, text width=7cm, draw=black]
\tikzstyle{choose} = [diamond, inner sep=1pt, minimum width=2cm, minimum height=2cm, text centered, text width=1.5cm, draw=black]
\tikzstyle{arrow} =[thick,->, >=stealth]

\pgfmathdeclarefunction{peierlspotential}{6}%
{
	\pgfmathparse
	{
		#2 / (#3 * #5) 
		* sin(deg(#1 * #3 - #3 * #5 * (x)))
		* exp(-( ( #1 - #5 * ((x) - #4) ) )^2.0 / ( #6^2.0 ) )
	}%
}

\pgfmathdeclarefunction{linearFct}{2}%
{%
	\pgfmathparse{#1*x+#2}%
}
\pgfmathdeclarefunction{quadFct}{3}%
{%
	\pgfmathparse{#1*x*x+#2*x+#3}%
}
\pgfmathdeclarefunction{logFct}{2}%
{%
	\pgfmathparse{#1*log10(x)+#2}%
}
\pgfmathdeclarefunction{expFct}{2}%
{%
	\pgfmathparse{#1*exp(-x/#2)}%
}

\newboolean{buildCurrentBlockInline}
\setboolean{buildCurrentBlockInline}{false}
\newboolean{rebuildCurrentBlock}
\setboolean{rebuildCurrentBlock}{false}

\newboolean{rebuildData}
\setboolean{rebuildData}{false}

\newboolean{removeData}
\setboolean{removeData}{false}

\graphicspath{{figures/}}

\newboolean{buildtikzpics}
\setboolean{buildtikzpics}{false}
\newif\ifrebuildtikz
\newif\ifChangeMode
\ChangeModetrue
\ChangeModefalse
\ifthenelse{\boolean{buildtikzpics}}
{
	\rebuildtikztrue
	\usetikzlibrary{external}
	\tikzexternalize[optimize=false,prefix=figures/autogen/]%
}
{
	\rebuildtikzfalse
}

\usepackage{scalerel}

\definecolor{orcidlogocol}{HTML}{A6CE39}
\tikzset{
	orcidlogo/.pic={
		\fill[orcidlogocol] svg{M256,128c0,70.7-57.3,128-128,128C57.3,256,0,198.7,0,128C0,57.3,57.3,0,128,0C198.7,0,256,57.3,256,128z};
		\fill[white] svg{M86.3,186.2H70.9V79.1h15.4v48.4V186.2z}
		svg{M108.9,79.1h41.6c39.6,0,57,28.3,57,53.6c0,27.5-21.5,53.6-56.8,53.6h-41.8V79.1z M124.3,172.4h24.5c34.9,0,42.9-26.5,42.9-39.7c0-21.5-13.7-39.7-43.7-39.7h-23.7V172.4z}
		svg{M88.7,56.8c0,5.5-4.5,10.1-10.1,10.1c-5.6,0-10.1-4.6-10.1-10.1c0-5.6,4.5-10.1,10.1-10.1C84.2,46.7,88.7,51.3,88.7,56.8z};
	}
}

\newcommand\orcidicon[1]{\href{https://orcid.org/#1}{\mbox{\scalerel*{
				\begin{tikzpicture}[yscale=-1,transform shape]
					\pic{orcidlogo};
				\end{tikzpicture}
			}{|}}}}

\usepackage{ulem}
\usepackage[colorlinks,bookmarks=false,citecolor=blue,linkcolor=black,urlcolor=blue]{hyperref}
\usepackage{cleveref}
\Crefname{appendix}{Appendix}{Appendices}
\Crefname{equation}{Equation}{Equations}
\Crefname{figure}{Figure}{Figures}
\Crefname{section}{Section}{Sections}
\Crefname{tabular}{Tabular}{Tabulars}
\crefname{appendix}{App.}{Apps.}
\crefname{equation}{Eq.}{Eqs.}
\crefname{figure}{Fig.}{Figs.}
\crefname{section}{Sec.}{Secs.}
\crefname{tabular}{Tab.}{Tabs.}

\ExplSyntaxOn
\DeclareExpandableDocumentCommand \eval { m } { \fp_eval:n { #1 } }
\ExplSyntaxOff

\makeatletter
\def\pgfplotsutil@decstringcounter#1{%
 \begingroup
  \c@pgf@counta=#1\relax
  \advance\c@pgf@counta by -1
  \edef#1{\the\c@pgf@counta}%
  \pgfmath@smuggleone#1%
 \endgroup
}%

\pgfplotsset{
/pgfplots/each nth point*/.style 2 args={%
/pgfplots/x filter/.append code={%
 \ifnum\coordindex=0
  \def\c@pgfplots@eachnthpoint@xfilter{0}%
  \edef\c@pgfplots@eachnthpoint@xfilter@cmp{#1}%
 \else
  \ifnum\coordindex>#2\relax
   \pgfplotsutil@advancestringcounter\c@pgfplots@eachnthpoint@xfilter
   \ifx\c@pgfplots@eachnthpoint@xfilter@cmp\c@pgfplots@eachnthpoint@xfilter
    \def\c@pgfplots@eachnthpoint@xfilter{0}%
   \else
    \let\pgfmathresult\pgfutil@empty
   \fi
  \fi
 \fi
}%
},
}
\makeatother

\newcommand{\printpgfnumberorder}[1]%
{%
	\pgfmathfloatparsenumber{#1}%
	\pgfmathfloattomacro{\pgfmathresult}{\Fn}{\Mn}{\En}%
	\pgfmathparse{\Fn==2 ? "-" : ""}%
	\edef\Sn{\pgfmathresult}%
	\Sn 10^{\En}%
}

\newcounter{marknumber}
\pgfplotsset{
	error bars/every nth mark/.style={
		/pgfplots/error bars/draw error bar/.prefix code={
			\pgfmathtruncatemacro\marknumbercheck{mod(floor(\themarknumber/2),#1)}
			\ifnum\marknumbercheck=0
			\else
			\begin{scope}[opacity=0]
				\fi
			},
			/pgfplots/error bars/draw error bar/.append code={
				\ifnum\marknumbercheck=0
				\else
			\end{scope}
			\fi
			\stepcounter{marknumber}    
		}
	}
}
\newacronym[shortplural={MPS}]{MPS}{MPS}{matrix\hyp product state}
\newacronym{MPO}{MPO}{matrix-product operator}
\newacronym{SVD}{SVD}{singular-value decomposition}
\newacronym{QCS}{QCS}{quantum-computer simulator}
\newacronym{FSM}{FSM}{finite-state machine}
\newacronym{ACA}{ACA}{adaptive cross approximation}
\newacronym{1D}{1D}{one\hyp dimensional}
\newacronym{QC}{QC}{quantum computer}
\newacronym{CDW}{CDW}{charge\hyp density wave}
\newacronym{BOW}{BOW}{bond\hyp order wave}
\newacronym{SDW}{SDW}{spin\hyp density wave}
\newacronym{ARPES}{ARPES}{angle-resolved photoemission spectroscopy}
\newacronym{OBC}{OBC}{open-boundary conditions}
\newacronym{PBC}{PBC}{periodic-boundary conditions}
\newacronym{TEBD}{TEBD}{time-evolution block-decimation}
\newacronym{TDVP}{TDVP}{time\hyp dependent variational principle}
\newacronym{iff}{iff}{if and only if}
\newacronym{DFT}{DFT}{density\hyp functional theory}
\newacronym{DMFT}{DMFT}{dynamical mean\hyp field theory}
\newacronym{DMRG}{DMRG}{density\hyp matrix renormalization group}
\newacronym{1DMRG}{1DMRG}{single-site density\hyp matrix renormalization group}
\newacronym{2DMRG}{2DMRG}{two-site density\hyp matrix renormalization group}
\newacronym{DMRG3S}{DMRG3S}{strictly single-site density\hyp matrix renormalization group}
\newacronym{iDMRG}{iDMRG}{inifinite\hyp size density\hyp matrix renormalization group}
\newacronym{tDMRG}{tDMRG}{time\hyp dependent density\hyp matrix renormalization group}
\newacronym{PP-DMRG}{PP-DMRG}{projected purified density\hyp matrix renormalization group}
\newacronym{QMC}{QMC}{quantum Monte Carlo}
\newacronym{AIM}{AIM}{Anderson impurity model}
\newacronym{SIAM}{SIAM}{single impurity Anderson model}
\newacronym{LDA}{LDA}{local\hyp density approximation}
\newacronym{LBNL}{LBNL}{Lawrence Berkeley National Laboratory}
\newacronym{VQE}{VQE}{variational\hyp quantum eigensolver}
\newacronym{ED}{ED}{exact diagonalization}
\newacronym{QPT}{QPT}{quantum phase transition}
\newacronym{QCP}{QCP}{quantum critical point}
\newacronym{ETH}{ETH}{eigenstate thermalization hypothesis}
\newacronym{EHM}{EHM}{extended Hubbard model}
\newacronym{AKLT}{AKLT}{Affleck\hyp Lieb\hyp Kennedy\hyp Tasaki}
\newglossaryentry{QR}{name={QR},description={QR decomposition}}
\newacronym{TNS}{TNS}{tensor\hyp network state}
\newacronym{SM}{SM}{supplemental material}
\newacronym{NOO}{NOO}{natural orbital occupation}
\newacronym{NO}{NO}{natural orbital}
\newacronym{LRO}{LRO}{long\hyp range order}
\newacronym{qLRO}{qLRO}{quasi\hyp long\hyp range order}
\newacronym{SC}{SC}{Superconductivity}
\newacronym{VBGS}{VBGS}{valence bond ground-state}
\newacronym{PEPS}{PEPS}{projected entangled pair\hyp states}
\newacronym{ALS}{ALS}{alternating least squares}
\newacronym{BdG}{BdG}{Bogoljubov de-Gennes}
\newacronym{TFIM}{TFI}{transverse field Ising model}
\newacronym{PP}{PP}{projected purification}
\newacronym{BEC}{BEC}{Bose\hyp Einstein condensate}
\newacronym{JWT}{JWT}{Jordan\hyp Wigner transformation}
\newacronym{NISQ}{NISQ}{noisy intermediate scale quantum}
\newacronym{NN}{NN}{nearest\hyp neighbor}
\newacronym{NNN}{NNN}{next\hyp nearest\hyp neighbor}
\newacronym{SPDM}{SPDM}{single\hyp particle density matrix} 
\newacronym{HCB}{HCB}{hardcore bosons}
\newacronym{SF}{SF}{spinless fermions}
\newacronym{fRG}{fRG}{functional renormalization group}
\newacronym{LE}{LE}{Luther\hyp Emery}
\newacronym{FQH}{FQH}{fractional quantum Hall}
\newacronym{FCI}{FCI}{fractional Chern insulators}
\newacronym{ODLRO}{ODLRO}{off-diagonal long-range order}
\newacronym{HODLRO}{HODLRO}{hidden off-diagonal long-range order}
\newcommand{\acsaddress}{Department of Physics and Arnold Sommerfeld Center for Theoretical Physics (ASC), Ludwig-Maximilians-Universit\"{a}t M\"{u}nchen, D-80333 Munich, Germany}
\newcommand{\mcqstaddress}{Munich Center for Quantum Science and Technology (MCQST), D-80799 M\"{u}nchen, Germany}
\newcommand{\harvardaddress}{Department of Physics, Harvard University, Cambridge, MA 02138, USA}
\newcommand{\itampaddress}{ITAMP, Harvard-Smithsonian Center for Astrophysics, Cambridge, MA 02138, USA}
\newcommand{\regensburgaddress}{Institute of Theoretical Physics, University of Regensburg, D-93053, Germany}

\newcommand{\nodagger}[0]{{\vphantom{\dagger}}}

\Crefname{appendix}{Appendix}{Appendices}
\Crefname{equation}{Equation}{Equations}
\Crefname{figure}{Figure}{Figures}
\Crefname{section}{Section}{Sections}
\Crefname{tabular}{Tabular}{Tabulars}
\crefname{appendix}{App.}{Apps.}
\crefname{equation}{Eq.}{Eqs.}
\crefname{figure}{Fig.}{Figs.}
\crefname{section}{Sec.}{Secs.}
\crefname{tabular}{Tab.}{Tabs.}
\begin{document}
\title{Detecting Hidden Order in Fractional Chern Insulators}
\author{F.~J.~Pauw* \orcidicon{0009-0006-4188-8503}}
\affiliation{\acsaddress}
\affiliation{\mcqstaddress}
\altaffiliation{These authors contributed equally to this work.}
\author{F.~A.~Palm* \orcidicon{0000-0001-5774-5546}}
\affiliation{\acsaddress}
\affiliation{\mcqstaddress}
\author{U.~Schollwöck \orcidicon{0000-0002-2538-1802}}
\affiliation{\acsaddress}
\affiliation{\mcqstaddress}
\author{A.~Bohrdt \orcidicon{0000-0002-3339-5200}}
\affiliation{\mcqstaddress}
\affiliation{\harvardaddress}
\affiliation{\itampaddress}
\affiliation{\regensburgaddress}
\author{S.~Paeckel \orcidicon{0000-0001-8107-069X}}
\affiliation{\acsaddress}
\affiliation{\mcqstaddress}
\author{F.~Grusdt \orcidicon{0000-0003-3531-8089}}
\affiliation{\acsaddress}
\affiliation{\mcqstaddress}
\date{\today}
\begin{abstract}
Topological phase transitions go beyond Ginzburg and Landau's paradigm of spontaneous symmetry breaking and occur without an associated local order parameter. Instead, such transitions can be characterized by the emergence of non-local order parameters, which require measurements on extensively many particles simultaneously - an impossible venture in real materials. On the other hand, quantum simulators have demonstrated such measurements, making them prime candidates for an experimental confirmation of non-local topological order. Here, building upon the recent advances in preparing few-particle fractional Chern insulators using ultracold atoms and photons, we propose a realistic scheme for detecting the hidden off-diagonal long-range order (HODLRO) characterizing Laughlin states. Furthermore, we demonstrate the existence of this hidden order in fractional Chern insulators, specifically for the $\nu=\nicefrac{1}{2}$-Laughlin state in the isotropic Hofstadter-Bose-Hubbard model. This is achieved by large-scale numerical density matrix renormalization group (DMRG) simulations based on matrix product states, for which we formulate an efficient sampling procedure providing direct access to HODLRO in close analogy to the proposed experimental scheme. We confirm the characteristic power-law scaling of HODLRO, with an exponent $\nicefrac{1}{\nu} = 2$, and show that its detection requires only a few thousand snapshots. This makes our scheme realistically achievable with current technology and paves the way for further analysis of non-local topological orders, e.g. in topological states with non-Abelian anyonic excitations.
\end{abstract}
\maketitle
\section{Introduction\label{sec:introduction}}
Over the last decades, interacting topological systems have provided an exciting opportunity to explore exotic states of matter, most prominently states exhibiting intrinsic topological order.
A prime example for this paradigm is the~\gls{FQH} effect, where strongly interacting particles in two dimensions are subject to a strong magnetic field~\cite{Tsui1982, Jain2007}.
In contrast to symmetry breaking phases of matter, such topologically ordered states cannot be described by a local order parameter.
Nevertheless, since the early days of FQH physics researchers have been intrigued by the idea of using non-local extensions of conventional order parameters in order to characterize topological states of matter and enable insights into their remarkable properties~\citep{Girvin1987,Girvin1988,Rezayi1988,Read1989}.
Specifically, based on variational trial states like the Laughlin state~\cite{Laughlin1983}, the emergence of so-called~\gls{HODLRO} has been established theoretically in continuum systems.
This multi-particle variant of conventional~\gls{ODLRO}, characterizing Bose-Einstein condensation associated with a global $\mathrm{U}(1)$-symmetry breaking, will play a central topic in this article.
As we will discuss, in order to detect HODLRO in a FQH system a measurement of long-range one-particle coherence, $\sim \hat{\psi}^{\dagger}(\boldsymbol{r}_1+\boldsymbol{d})\hat{\psi}^{\nodagger}(\boldsymbol{r}_1)$, has to be combined with a simultaneous measurement of the positions $\boldsymbol{r}_2,...,\boldsymbol{r}_N$ of all remaining $N-1$ particles.
This makes a direct measurement of HODLRO unfeasible in traditional solid state experiments.
\par
In contrast, modern quantum simulator platforms, such as ultracold atoms in quantum gas microscopes~\cite{Bakr2010,Sherson2010,Gross2021} or arrays of superconducting qubits~\cite{Wilkinson2020,Kjaergaard2021}, routinely perform such measurements of all particles simultaneously, achieving full single-site and single-particle resolution.
Furthermore, these quantum simulators have recently begun to explore the interplay of topological bandstructures and strong interactions.
Using lattice shaking or Raman transitions, artificial gauge-fields have been implemented in optical lattices by realizing complex
tunneling matrix elements for the neutral atoms.
\begin{figure*}[ht]
	\centering
	\includegraphics{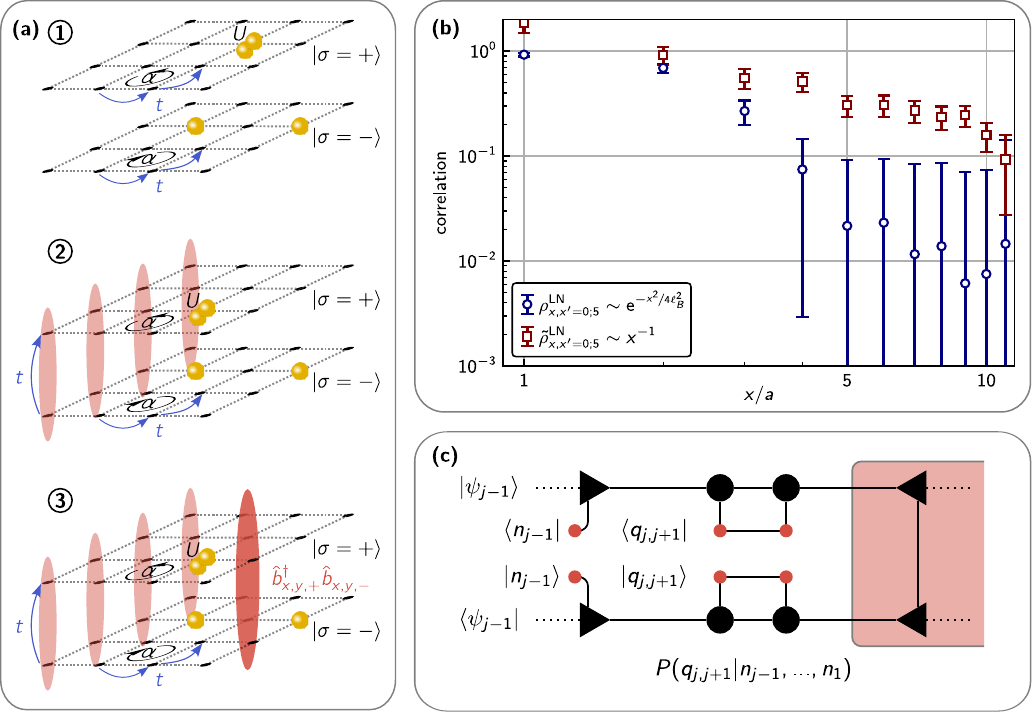}
	\caption{
		\textbf{(a)}  \raisebox{.5pt}{\textcircled{\raisebox{-.9pt} {\textbf{1}}}} Possible experimental realization of the proposed measurement protocol using a synthetic bilayer Hofstadter-Hubbard system with internal states $\ket{\sigma=\pm}$.
		\raisebox{.5pt}{\textcircled{\raisebox{-.9pt} {\textbf{2}}}} On one side of the system, the edges are coupled to obtain an effective single layer system.
		 \raisebox{.5pt}{\textcircled{\raisebox{-.9pt} {\textbf{3}}}} Measurements of off-diagonal elements of the one-particle reduced density matrix, $\left\langle\hat{a}_{x,y,+}^{\dagger} \hat{a}_{x,y,-}^{\phantom\dagger}\right\rangle$ are possible using a local pulse protocol involving both layers, as described in the main text.
		\textbf{(b)} Correlations extracted from $N_{\rm snaps}=10^4$ numerical snapshots without any post-processing and with experimentally relevant errorbars from shot-noise.
		The algebraic decay of the two-site correlations of the composite bosons $|\tilde{\rho}_{x, x^{\prime}; y}| \sim \left\vert \left\langle\hat{b}_{x,y}^{\dagger} \hat{b}_{x^{\prime},y}\right\rangle \right\vert$ (red squares) can be clearly distinguished from the exponential decay of the original bosonic correlations $|\rho_{x,x^{\prime}; y}| \sim \left\vert \braket{\hat{a}_{x,y}^{\dagger} \hat{a}_{x^{\prime}, y}}\right\vert$ (blue circles).
		\textbf{(c)} Tensor-network diagram of the extended perfect sampling scheme which allows for numerical snapshots of composite boson correlations $|\tilde{\rho}_{x, x^{\prime}; y}|$ equivalent to the ones obtained in the proposed experiment.  
	}
	\label{fig:Sketch}
\end{figure*}
In particular, implementations of artificial gauge fields~\cite{Aidelsburger2013,Miyake2013} along with strong interactions~\cite{Tai2017} allowed for a first realization of a two-particle $\nu=\nicefrac{1}{2}$-Laughlin state in an optical lattice~\cite{Leonard2023} based on the Hofstadter-Bose-Hubbard Hamiltonian \cite{Soerensen2005,Hafezi2007,Palmer2008,Moeller2009,Moeller2015,Bauer2016,Huegel2017,Motruk2017,He2017,Gerster2017,Andrews2018,Dong2018,Andrews2021,Palm2021,Andrews2021a,Boesl2022,Wang2022,Palm2022}.
Similarly, experiments have been successful at creating a Laughlin state made of light in a photonic analog of a FQH system~\cite{Clark2020}.
Other experimental angles such as microwave cavity arrays are also promising candidates to engineer topological materials~\cite{Anderson2016}.
\par
In this article we show that state-of-the-art quantum simulation platforms, and cold atoms in particular, can be used to directly detect HODLRO in FQH states.
Through large-scale numerical~\gls{DMRG} simulations \cite{White1992,Schollwock2011,Hubig2015} of the isotropic Hofstadter-Bose-Hubbard model at magnetic filling $\nu=\nicefrac{1}{2}$, we demonstrate that fractional Chern insulators in lattice systems exhibit the same form of universal HODLRO as the corresponding continuum FQH states.
To this end, we perform projective measurements on the~\gls{MPS} representation of the ground state, building on previous works introducing a perfect sampling algorithm for~\gls{MPS}~\cite{Ferris2012} to emulate snapshots of local observables~\cite{Buser2022} via a successive collapse of the many\hyp body wave\hyp function.
We extend this scheme in such a way that it enables snapshots of non\hyp local quantities (see Fig.~\ref{fig:Sketch}(c)) to sample correlation functions like those needed to detect the multi-particle HODLRO.
Our numerical prediction can be experimentally tested using two hyperfine manifolds in ultracold atoms in a quantum gas microscope (see Fig.~\ref{fig:Sketch}(a)), for which we propose a specific experimental implementation enabling long-range measurements of the one-particle coherence combined with simultaneous single-site resolved measurements of the density on the whole lattice, granting direct access to HODLRO (see Fig.~\ref{fig:Sketch}(b)).
\par
The remainder of this paper is structured as follows:
In Section \ref{sec:proposal} we present an overview of the main results of this article. 
We briefly introduce the key concepts of HODLRO and elucidate explicitly how to obtain snapshots necessary for the detection of this putative multi-particle off-diagonal long-range order from measurements on cold atoms in optical lattices.
In Section \ref{sec:Numerical Analysis} we extensively discuss the results of our large-scale numerical analysis of the Hofstadter-Bose-Hubbard ground state showing the absence of conventional long-range order and the emergence of HODLRO using the MPS-based sampling method.
Finally, in Section \ref{sec:methods} we present the MPS-based projective sampling scheme in a self-contained review which can be read independently.

\section{Overview of Main results \label{sec:proposal}}
This section summarizes the main findings of the paper.
We begin with a short review of HODLRO in continuum FQH systems.
We then proceed by generalizing this concept to lattice systems hosting  fractional Chern insulators.
After introducing a scheme to resolve long-range coherence in quantum gas microscopes, exploiting a bilayer geometry, a detailed experimental setup is proposed which yields all information needed to obtain HODLRO.
\subsection{A brief Review of HODLRO}
The ordering in superfluids and superconductors can be famously characterized by the existence of off-diagonal long-range order (ODLRO) associated with the breaking of the particle number preserving $\mathrm{U}(1)$ gauge symmetry of the system signaling Bose-Einstein condensation. 
Consider the one-particle reduced density matrix
\begin{equation}
	\rho_{\boldsymbol{r},\boldsymbol{r}+\boldsymbol{d}} = \langle \hat{\psi}^{\dagger}(\boldsymbol{r})\hat{\psi}^{\nodagger}(\boldsymbol{r}+\boldsymbol{d}) \rangle.
\end{equation}
In general, one expects $\rho_{\boldsymbol{r},\boldsymbol{r}+\boldsymbol{d}}$ to fall off rapidly and decay to zero for increasing values of $\boldsymbol{d}$.
In systems exhibiting ODLRO  however, the one-particle correlations saturate to a finite value in this limit
\begin{equation}
	\lim_{\boldsymbol{d}\rightarrow \infty} \rho_{\boldsymbol{r},\boldsymbol{r}+\boldsymbol{d}} \neq 0.
\end{equation}
\par
Contrary, one of the hallmarks of topologically ordered states of matter is the absence of any kind of one-particle long-range order. 
For example, for Laughlin's trial wave function describing the ground state of a continuum FQH system at magnetic filling $\nu=\nicefrac{1}{m}$,
\begin{equation}
	\psi_{\rm LN}(z_1,...,z_N)= \prod_{j<k} (z_j-z_k)^{m} \mathrm{e}^{-\frac{1}{4}\sum_{l}|z_l|^2},
\end{equation}
where $z_j=x_j+iz_j$ is the $j\rm th$ particle position in the complex plane, it can be shown that the one-particle reduced density matrix decays exponentially
\begin{equation}
	\rho^{\rm LN}_{z,z^{\prime}} = \langle \hat{\psi}^{\dagger}(z)\hat{\psi}^{\nodagger}(z^{\prime}) \rangle \sim \mathrm{e}^{-\frac{1}{4} (z-z^{\prime})^2/\ell_B^2}.
\end{equation}
\par
Nonetheless, there is a characteristic type of order hidden in the density matrix of the Laughlin state.
It can be revealed by introducing a singular gauge transformation~\cite{Wilczek1982,Arovas1984,Arovas1985} of the form
\begin{equation} \label{Eq:SGT}
	\mathcal{A}(z_1,...,z_N) = \frac{m \Phi_0}{2\pi}\sum_j\sum_{k\neq j} \nabla_j \mathrm{Im}\{\mathrm{ln}(z_k-z_j)\}.
\end{equation}
The effect of this transformation is a contribution of the form $\prod_{j<k} \nicefrac{\vert z_j - z_k \vert^m}{(z_j-z_k)^m}$ to the wave function which can be understood as endowing each of the particles with $m$ quanta of magnetic flux $\Phi_0$.
Independent of $m$ being odd/even, Eq.~\eqref{Eq:SGT} maps the fermionic/bosonic FQH problem to composites of flux quanta and charge with bosonic statistics which are hence referred to as \textit{composite bosons} $\hat{\psi}_{CB}^{(\dagger)}(z_j)$.
For a more detailed derivation, see Appendix~\ref{App:SGT}.
In this composite boson basis the continuum one-particle reduced density matrix decays algebraically, i.e.
\begin{equation}
	\tilde{\rho}^{\rm LN}_{z,z^{\prime}} = \langle \hat{\psi}_{CB}^{\dagger}(z)\hat{\psi}_{CB}^{\nodagger}(z^{\prime}) \rangle  \sim |z-z^{\prime}|^{-2\nu}.
\end{equation}
This emergence of quasi long-range order, signaling a condensation of the composite bosons in the Laughlin state~\cite{Girvin1988,Read1989} defines \textit{hidden} off-diagonal long range order (HODLRO).	
\subsection{HODLRO in Lattice Systems}
We can readily expand the concept of HODLRO to discrete systems.
Inspired by earlier studies of lattice anyons~\cite{Fradkin1989,Fradkin1990}, in a lattice system with local bosonic degrees of freedom $\hat{a}^{(\dagger)}_{\boldsymbol{k}}$, attaching $m$ flux quanta to each boson can be achieved by the following lattice gauge transformation
\begin{equation} \label{CB}
	\hat{\Phi}_{\boldsymbol{j}} = m \sum_{\boldsymbol{k} \neq \boldsymbol{j}} \Theta(\boldsymbol{j},\boldsymbol{k}) \hat{n}_{\boldsymbol{k}} , \quad \Theta (\boldsymbol{j},\boldsymbol{k}) = \arg(z_{\boldsymbol{j}} - z_{\boldsymbol{k}})
\end{equation}
where the bold letters represent the coordinates of a particular lattice site, e.g. $\boldsymbol{k}= (x,y)$ and $\hat{n}_{\boldsymbol{k}}=\hat{a}^{\dagger}_{\boldsymbol{k}}\hat{a}^{\nodagger}_{\boldsymbol{k}}$ is the local density operator.
Evidently, $\hat{\Phi}_{\boldsymbol{j}}$ is the lattice analog of the continuum emergent gauge degrees of freedom of $\mathcal{A}_j$.
Hence, using Eq.~\eqref{CB}, we can define composite boson operators
\begin{equation} \label{CB operator}
	\hat{b}_{\boldsymbol{j}}^{(\dagger)} := e^{(-)i \hat{\Phi}^{(\dagger)}_{\boldsymbol{j}}} \hat{a}_{\boldsymbol{j}}^{(\dagger)}, 
\end{equation}
describing the desired composites of $m$ flux quanta and a boson.
\par
In turn, we can introduce the one-particle correlation function for the composite bosons,
\begin{equation}
	\tilde{\rho}_{\boldsymbol{j},\boldsymbol{l}} = \left\langle \hat{b}_{\boldsymbol{j}}^{\dagger} \hat{b}_{\boldsymbol{l}}^{\nodagger} \right\rangle,
\end{equation}
which can be expressed purely in terms of the original bosonic annihilation (creation) operators $\hat{a}_{\boldsymbol{j}}^{(\dagger)}$ and the boson number operators $\hat{n}_{\boldsymbol{k}}$
\begin{equation} \label{Eq:LatticeHODLRO}
	\tilde{\rho}_{\boldsymbol{j},\boldsymbol{l}} = \left\langle \prod_{\boldsymbol{k} \neq \boldsymbol{j},\boldsymbol{l}} \left( \frac{z_{\boldsymbol{j}} - z_{\boldsymbol{k}} }{|z_{\boldsymbol{j}} - z_{\boldsymbol{k}}|} \right)^{-m \hat{n}_{\boldsymbol{k}}} \left( \frac{z_{\boldsymbol{l}} - z_{\boldsymbol{k}} }{|z_{\boldsymbol{l}} - z_{\boldsymbol{k}}|} \right)^{m \hat{n}_{\boldsymbol{k}}} \hat{a}^{\dagger}_{\boldsymbol{j}} \hat{a}^{\nodagger}_{\boldsymbol{l}} \right\rangle.
\end{equation}
Consequently, the multi-particle composite boson correlation function $\tilde{\rho}_{\boldsymbol{j},\boldsymbol{l}}$ picks up an additional non-trivial phase for each occupied lattice site $\boldsymbol{k} \neq \boldsymbol{j},\boldsymbol{l}$ with respect to the usual correlator $\rho_{\boldsymbol{j},\boldsymbol{l}} \sim \langle \hat{a}_{\boldsymbol{j}}^{\dagger}\hat{a}_{\boldsymbol{l}}^{\nodagger} \rangle$. 
Therefore, in order to probe this lattice analog of HODLRO one not only has to measure the long-range bosonic one-particle coherence $\rho_{\boldsymbol{j},\boldsymbol{l}}$  but combine it with a simultaneous measurement of local densities operators $\{\hat{n}_{\boldsymbol{k}}\}_{\boldsymbol{k}\neq \boldsymbol{j},\boldsymbol{l}}$.
This enables us to examine lattice models described by the Hofstadter-Bose-Hubbard Hamiltonian
\begin{equation}
	\hat{\mathcal{H}} = -t\sum_{\langle \boldsymbol{j},\boldsymbol{k} \rangle} \mathrm{e}^{i\phi_{\boldsymbol{j}\rightarrow \boldsymbol{k}}}\hat{a}^{\dagger}_{\boldsymbol{j}}\hat{a}^{\nodagger}_{\boldsymbol{k}} +  \rm{H.c.}
\end{equation}
which hosts fractional Chern insulators.
The phase $\phi_{\boldsymbol{j}\rightarrow \boldsymbol{k}}$  is determined by the choice of the magnetic vector potential and the sum runs over adjacent sites. 
\subsection{Measuring Off-diagonal Long-range Order} \label{subsec:bilayer}
Despite being highly adjustable and offering single-lattice-site and single-particle resolution, so far quantum-gas microscopes only allow for a direct measurement of physical quantities which are diagonal or near-diagonal in the occupation number basis.
To resolve phase coherence one relies on a combination of a time-of-flight expansion followed by absorption imaging granting access to the momentum distribution~\cite{Greiner2002}.
That means, even though there has been recent progress in resolving coherence and currents between adjacent sites~\cite{Killi2012,Kessler2014}, measuring long-range single-particle correlation functions of the form
\begin{equation}
	\rho_{\boldsymbol{r},\boldsymbol{r^{\prime}}} = \langle \hat{a}_{\boldsymbol{r}}^{\dagger}\hat{a}^{\nodagger}_{\boldsymbol{r^{\prime}}} \rangle , \quad |\boldsymbol{r}-\boldsymbol{r^{\prime}}|>1
\end{equation}
from single-site measurements is a non-trivial task. 
\par
Here, we propose to use a bilayer geometry consisting of two hyperfine manifolds of ultracold atoms which are coupled along one edge of the layers, see Fig.~\ref{fig:Sketch}(a). 
Using such a setup an effective single-layer is achieved in which we can realize a two-dimensional lattice model.
Exploiting this double layer structure, non\hyp local inter-lattice-site correlations $\sim \hat{a}^{\dagger}_{\boldsymbol{r}}\hat{a}^{\nodagger}_{\boldsymbol{r^{\prime}}}$ of the effective model, where $\boldsymbol{r}$ and $\boldsymbol{r^{\prime}}$ denote sites in different layers but are on top of each other - i.e. $\boldsymbol{r}=(x,y,+), \boldsymbol{r^{\prime}}=(x,y,-)$ -, can be accessed by measuring local correlations between the layers.  
In order to extract this inter-layer correlations, we propose to perform a local $\nicefrac{\pi}{2}$-pulse on the probe sites $\boldsymbol{r}$ and $\boldsymbol{r^{\prime}}$ to rotate the measurement basis into a coherent superposition of both layers. 
In this manner, the inter-layer correlations can be extracted using standard single-site resolved quantum gas microscopy.
In the following subsection, we will discuss in detail how such measurements can be performed and how the bilayer structure enables us to naturally treat bosonic atoms in an artificial, homogeneous magnetic field, making it a prime candidate to probe fractional Chern insulators.  
\subsection{Experimental Protocol}
In order to detect HODLRO in a cold atom quantum simulator we first need to achieve a long-range measurement of one-particle coherence and secondly, perform projective measurements on the site-local Fock spaces of the remaining sites simultaneously. 
For a particularly simple measurement of long-range two-point correlations $\sim \hat{a}_{\boldsymbol{r}}^{\dagger}\hat{a}_{\boldsymbol{r}^{\prime}}^{\nodagger}$, we exploit the bilayer approach described in Section~\ref{subsec:bilayer}.
In particular, we propose to realize a synthetic bilayer Hofstadter-Bose-Hubbard model using internal states $\ket{\pm}$ of the atoms to realize the different manifolds.
As already demonstrated experimentally~\cite{Aidelsburger2013,Miyake2013}, artificial gauge fields can be realized with opposite signs for the two internal states of the atom.
Thus, it is possible to realize a Hamiltonian of the form
\begin{equation} \label{Eq:ExpModel}
	\begin{aligned}
		\hat{\tilde{\mathcal{H}}}_{\rm exp} = &-t \sum_{x, y, \sigma} \left( \hat{a}_{x+1,y,\sigma}^{\dagger} \hat{a}_{x,y,\sigma}^{\nodagger}\right.\\
		&\phantom{-t \sum_{x, y, \sigma} \left(\right.}\left.+ \mathrm{e}^{\sigma 2\pi i\alpha (x+1/2)} \hat{a}_{x,y+1, \sigma}^{\dagger} \hat{a}_{x,y,\sigma}^{\nodagger} + \mathrm{H.c.} \right)\\
		& + \frac{U}{2} \sum_{x,y,\sigma} \hat{n}_{x,y,\sigma} \left(\hat{n}_{x,y,\sigma} - 1\right),
	\end{aligned}
\end{equation}
where $\hat{a}_{x,y,\sigma}^{\left(\dagger\right)}$ are the bosonic annihilation (creation) operators and $\hat{n}_{x,y,\sigma} = \hat{a}_{x,y,\sigma}^{\dagger} \hat{a}_{x,y,\sigma}^{\nodagger}$ are the boson number operators.
Here, $(x, y)$ are the lattice positions on a square lattice of size $L_x\times L_y$ parametrizing a single layer while $\sigma = \pm$ labels the internal state of the atoms. 
The Hamiltonian is written in Landau gauge, i.e. the physical vector potential which generates the homogeneous flux per plaquette $\alpha$ only affects the hopping amplitudes in $y$-direction which pick up a complex phase factor.
\par
While Eq.~\eqref{Eq:ExpModel} describes a true bilayer system, by locally driving the internal transition and hence introducing a local coupling between the internal states on the $x=0$ edge of the system, i.e. introducing a term
\begin{equation}
	\hat{\mathcal{H}}_{\rm exp} = \hat{\tilde{\mathcal{H}}}_{\rm exp} - t \sum_y \left(\hat{a}^{\dagger}_{0,y,+} \hat{a}^{\nodagger}_{0,y,-} + \mathrm{H.c.}\right),
\end{equation}
we can realize an effective single-layer model.
Note that the additional edge term only consists of hopping contributions in $x$ direction and thus, due to the gauge choice, does not contain complex phase factors. 
In this manner, an effective single-layer Hofstadter-Bose-Hubbard model of dimension $2L_x \times L_y$ with homogeneous artificial magnetic field of $\alpha$ flux quanta per plaquette can be realized, see Fig.~\ref{fig:Sketch}(a).\footnote{Another option is to realize a true spatial bilayer system and to couple the edge by driving a Raman transition.}
\par
The main advantage of such a synthetic bilayer system is that it allows for a direct measurement of $\hat{a}_{x,y,+}^{\dagger}\hat{a}_{x,y,-}^{\nodagger}$ by using local tunneling pulses between the layers as follows.
Switching on an additional coupling between the two internal states at site $(x,y,\pm)$ only and for a time corresponding to a $\nicefrac{\pi}{2}$-pulse essentially rotates the bosonic operators to a new basis 
\begin{equation}
	\hat{a}_{x,y}^{\pm} = \left(\hat{a}_{x,y,+} \pm \hat{a}_{x,y,-} \right)/\sqrt{2}.
\end{equation}
Now, applying a local detuning $\Delta$ between the internal states~\cite{Weitenberg:2011} for some time $\tau$ allows us to furthermore rotate the measurement basis to
\begin{equation}
	\hat{\tilde{a}}_{x,y}^{\pm} = \left(\hat{a}_{x,y,+} \pm i\mathrm{e}^{-i\varphi}\hat{a}_{x,y,-} \right)/\sqrt{2}
\end{equation}
where the relative phase $\varphi-\pi/2 \propto \Delta \tau$ can be controlled by the offset $\Delta$.
\par
Measuring the local occupation numbers with site and internal state resolution gives the density in the transformed basis $\hat{\tilde{n}}^{\pm}_{x,y}$ at the site where the pulse was applied, while at all other sites the original densities are recovered.
We note that in the original basis the transformed occupation numbers translate to
\begin{equation} \label{Eq:rotatedBasis}
	\hat{\tilde{n}}^{\pm}_{x,y} = \left(\hat{n}_{x,y,+} + \hat{n}_{x,y,-} \pm \hat{j}_{x,y}(\varphi)\right)/2,
\end{equation}
where we introduced the generalized current operator 
\begin{equation}
	\hat{j}_{x,y}(\varphi) = -i \mathrm{e}^{i\varphi} \hat{a}_{x,y,-}^{\dagger}\hat{a}_{x,y,+}^{\nodagger} + \mathrm{H.c.}.
\end{equation}
From Eq.~\eqref{Eq:rotatedBasis} it follows immediately that
\begin{equation}
	 \hat{\tilde{n}}_{x,y}^{+}+	\hat{\tilde{n}}_{x,y}^{-} = \hat{n}_{x,y,+} + \hat{n}_{x,y,-} \quad, \hat{\tilde{n}}_{x,y}^{+}-	\hat{\tilde{n}}_{x,y}^{-}=\hat{j}_{x,y}.
\end{equation}
That is, from occupation number measurements in the rotated basis, we can recover the occupation numbers in the initial basis, while simultaneously obtaining $\hat{j}_{x,y}(\varphi)$ for any value of the offset phase $\varphi$ resolving the absolute value of the coherence between the two sites via
\begin{equation} \label{Eq:Measurement}
	\vert \langle \hat{a}_{x,y,+}^{\dagger} \hat{a}_{x,y,-}^{\nodagger}\rangle \vert = \frac{1}{2}\sqrt{\langle \hat{j}_{x,y}(\nicefrac{\pi}{2})\rangle^2 + \langle\hat{j}_{x,y}(0)\rangle^2}
\end{equation}
\par
In general, exploiting the mapping of the bilayer system to an effective single-layer model, long-range correlations can be extracted using this protocol while reading out Fock basis snapshots on all remaining sites simultaneously.
Consequently, the protocol proposed here furthermore allows for measurements of more complex correlators of the form
\begin{equation}
	  \langle \hat{a}_{x,y,+}^{\dagger} \hat{a}_{x,y,-}^{\nodagger} f(\hat{n}_{x_2,y_2,\sigma_2}, \hdots, \hat{n}_{x_{N},y_{N},\sigma_{N}}) \rangle .
\end{equation}
In this case Eq.~\eqref{Eq:Measurement} generalizes to 
\begin{equation}
	\frac{1}{2}\sqrt{\langle \hat{j}_{x,y}(\nicefrac{\pi}{2}) f(\hat{n}_{x_j,y_j,\sigma_j};\nicefrac{\pi}{2}) \rangle^2 + \langle\hat{j}_{x,y}(0) f(\hat{n}_{x_j,y_j,\sigma_j};0) \rangle^2}
\end{equation}
which especially contains all the necessary information to compute HODLRO as defined in Eq.~\eqref{Eq:LatticeHODLRO} if we identify the function $f(\hat{n}_{x_j,y_j,\sigma_j})$ with the phase contribution due to the flux attachment.
Therefore, our protocol makes it feasible to extract this indicator of intrinsic topological order in state-of-the-art cold atom experiments.
\par
The remainder of this paper will numerically explore the possibility to use our approach for the bosonic Laughlin state at magnetic filling $\nu=\nicefrac{1}{2}$.
We find that already $N_{\rm{snaps}}=10^3$ numerical snapshots (see Fig.~\ref{fig:Sketch}(c)) of a system of $10\times10$ sites are sufficient to obtain not only a qualitative agreement with theoretical predictions from the continuum, but also an accurate quantitative estimate for the exponent of the power-law scaling of the HODLRO parameter (see Fig.~\ref{fig:Sketch}(b)), giving direct insights about the topological order of the probed quantum state.
\section{Numerical Analysis\label{sec:Numerical Analysis}}
In this section, we present an overview of the key results of the extensive numerical analysis of HODLRO we performed for the Hofstadter-Bose-Hubbard model.
First, we introduce the explicit model and the parameters considered in our simulations.
After pointing out key features of the ground states like incompressibility and the presence of a homogeneous density droplet in the bulk, we continue by confirming the absence of ordinary ODLRO.
Finally, we present strong indications for the emergence of HODLRO in the composite boson one-particle correlations of the obtained ground states.
\subsection{Model}
We study interacting bosons on an $L\times L$ square lattice  with open boundary conditions subject to a perpendicular magnetic field, modeled by the Hofstadter-Bose-Hubbard Hamiltonian
\begin{equation}\label{Eq:HBH}
\begin{aligned}
	\hat{\mathcal{H}} = &-t \sum_{x, y} \left( \hat{a}_{x+1,y}^{\dagger} \hat{a}_{x,y}^{\nodagger} + \mathrm{e}^{2\pi i\alpha x} \hat{a}_{x,y+1}^{\dagger} \hat{a}_{x,y}^{\nodagger} + \mathrm{H.c.} \right)\\
	& + \frac{U}{2} \sum_{x,y} \hat{n}_{x,y} \left(\hat{n}_{x,y} - 1\right),
\end{aligned}
\end{equation}
Here, $\hat{a}_{x,y}^{\left(\dagger\right)}$ are the bosonic annihilation (creation) operators and $\hat{n}_{x,y} = \hat{a}_{x,y}^{\dagger} \hat{a}_{x,y}^{\nodagger}$ are the boson number operators.
Analogously to the proposed experimental protocol, we perform our simulations using the Landau gauge.
However all results discussed below, including the correlations indicating HODLRO, also apply to other gauge choices, see Appendix \ref{app:D}.
We fix the flux per plaquette to $2\pi\alpha = \nicefrac{2\pi}{6}$ in all our simulations.
\begin{figure*}
	\centering
	\includegraphics{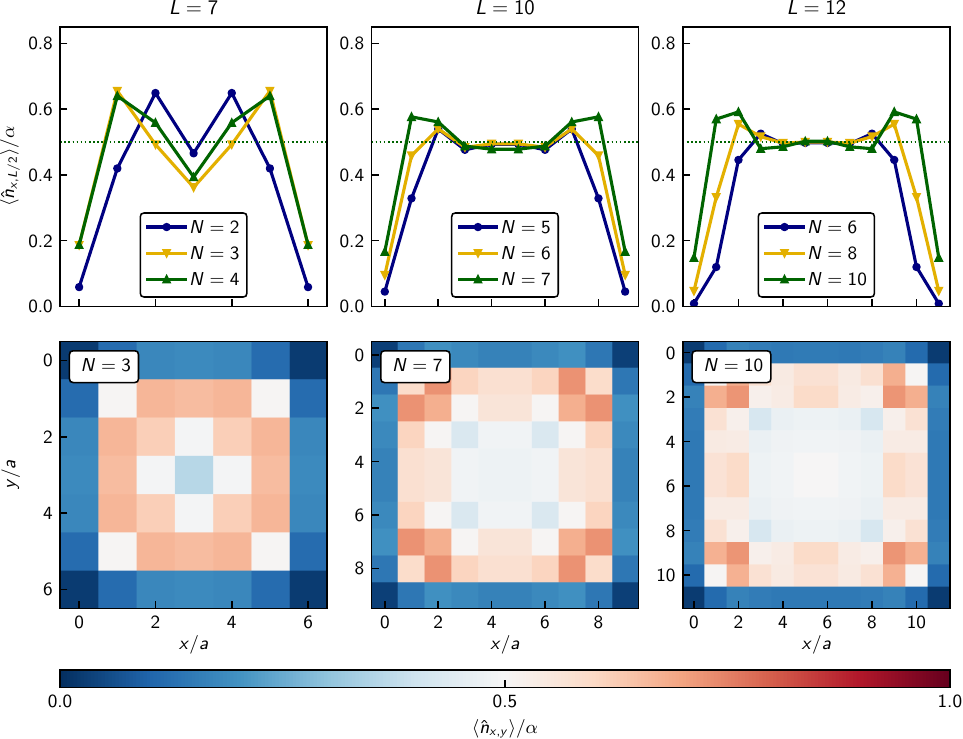}
	\caption{
		Local particle number density $\left\langle\hat{n}_{x,y}\right\rangle$ in units of the flux density $\alpha$ for different system sizes and particle numbers.
		We find an extended bulk region for sufficiently large systems ($L=10, 12$) exhibiting a bulk density consistent with the prediction $\nicefrac{\bar{n}}{\alpha}\approx\nicefrac{1}{2}$ for the Laughlin state (indicated by the dotted line in the upper row).
		 In the upper panel it is also demonstrated that these findings are robust to small changes in the particle number $N$, indicating the (bulk) incompressibility of the studied ground states.
	}
	\label{Fig:Density}
\end{figure*}
We study the hard-core bosonic limit, $\nicefrac{U}{t} \to \infty$, and consider dilute systems of few bosons $N$, with $\nicefrac{N}{L^2} \ll 1$.
Similar models have been explored in the literature for a while and it was established that the ground state of Eq.~\eqref{Eq:HBH} considering the set of parameters listed above close to magnetic filling factor $\nu=\nicefrac{N}{N_{\phi}} = \nicefrac{1}{2}$ is a lattice analog of the Laughlin state~\cite{Laughlin1983,Soerensen2005,Hafezi2007,Motruk2017,Gerster2017,Boesl2022,Wang2022,Palm2022,Leonard2023} where $N_{\phi}$ is the number of flux quanta piercing the sample.
\begin{figure*}[ht]
	\centering
	\includegraphics{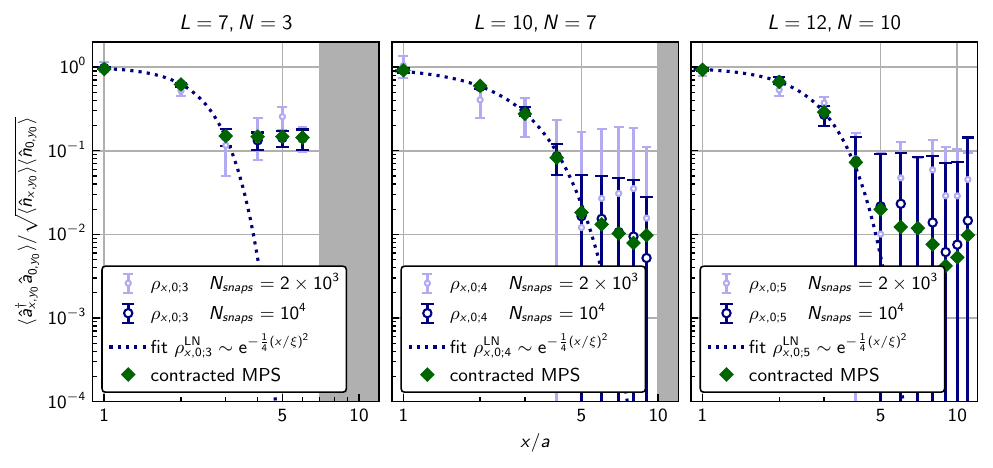}
	\caption{
		Normalized two-point correlation function for the system sizes and particle numbers of Fig.~\ref{Fig:Density} with the reference site on the 
		edge, i.e. $x^{\prime}=0$.
		While we observe significant finite size effects, we find the characteristic exponential decay in all systems using a snapshot based scheme (circles, $N_{\rm snaps}=2\times 10^{3}, 10^4$) with an exponential fit (dotted line) in good agreement.
		The errorbars of the snapshots correspond to a single standard deviation $\pm \nicefrac{\sigma_{\rho_{x,0;y}}}{\sqrt{N}}$.
		Furthermore, a direct contraction of the ground state MPS (diamonds) confirms these findings, in particular at short and intermediate length scales.
		The correlation length of the fitted decay is $\xi \simeq \nicefrac{3a}{2}$ which is significantly lager than the magnetic length $l_B \simeq a < \xi$.
	}
	\label{Fig:ExponentialDecay}
\end{figure*}
Furthermore, it has been shown that a finite but strong Hubbard repulsion $\nicefrac{U}{t}$ is already sufficient to stabilize the Laughlin state.
Therefore, we expect the results obtained in the hard-core limit $\nicefrac{U}{t}\to\infty$ to carry over to the experimentally relevant case of strong but finite interactions.
\\
\par

We use the single-site DMRG method~\cite{White1992,Schollwock2011,Hubig2015} implemented in the \textsc{SyTen}-toolkit~\cite{SyTen} to find an MPS representation of the ground state of Eq.~\eqref{Eq:HBH} with maximum bond dimension $\chi=1000$, where we exploit the $\mathrm{U}(1)$ symmetry associated with particle number conservation of the system.
We consider different system sizes ($L  = 7, 10, 12$) and different particle numbers $N$, however remaining in the dilute limit, $\nicefrac{N}{L^2} \ll 1$.
The magnetic length, and thus the correlation length of the Laughlin state, is given by $\ell_B=\nicefrac{a}{\sqrt{2\pi\alpha}} \approx a$, where $a$ is the lattice constant.

We calculate the local density $\left\langle\hat{n}_{x,y}\right\rangle$ for the DMRG ground states and, for sufficiently large systems ($L = 10, 12$), identify a bulk region of density \mbox{$\bar{n} \approx \nu\alpha = \nicefrac{\alpha}{2}$} as expected for a $\nu=\nicefrac{1}{2}$-Laughlin state, see Fig.~\ref{Fig:Density} (lower panel).
In particular, this behavior is robust for a range of particle numbers satisfying $\nicefrac{N}{L^2} \approx \nicefrac{\alpha}{2}$, which can be interpreted as a signature of the incompressibility of the Laughlin state, see Fig.~\ref{Fig:Density} (upper panel).
For the smallest systems studied here ($L=7$), the situation is less clear because of significant finite-size effects, and no extended bulk region is formed.
However, earlier studies found evidence of an approximate Laughlin state even in systems of $4\times 4$ sites~\cite{Leonard2023}, so that it might still be possible to find signatures of the topological nature in very small systems.
\subsection{Absence of ODLRO}
As discussed above, topologically ordered states of matter are prominently characterized by the absence of conventional long-range order, and in particular ODLRO.
We confirm this to be true for the states under study here by analyzing the behavior of the normalized two-point correlation function along the $x$-direction
\begin{equation}
	\rho_{x,x^{\prime}; y} =\left\langle \hat{a}_{x,y}^{\dagger} \hat{a}_{x^{\prime},y}\right\rangle/\sqrt{\left\langle \hat{n}_{x,y}\right\rangle \left\langle \hat{n}_{x^{\prime},y}\right\rangle}.
\end{equation}
Analytical continuum calculations for the Laughlin states at general filling factor $\nu = \nicefrac{1}{m}$ for a fixed value of $y$ found a characteristic exponential decay at long distances,
\begin{equation}\label{exp_continuum}
	\rho_{x,x^{\prime}; y}^{\rm LN} \sim \mathrm{e}^{-\frac{1}{4} (x-x^{\prime})^2/\ell_B^2},
\end{equation}
which is in particular independent of the filling factor~\cite{Girvin1988}.
\begin{figure*}[ht]
	\centering
	\includegraphics{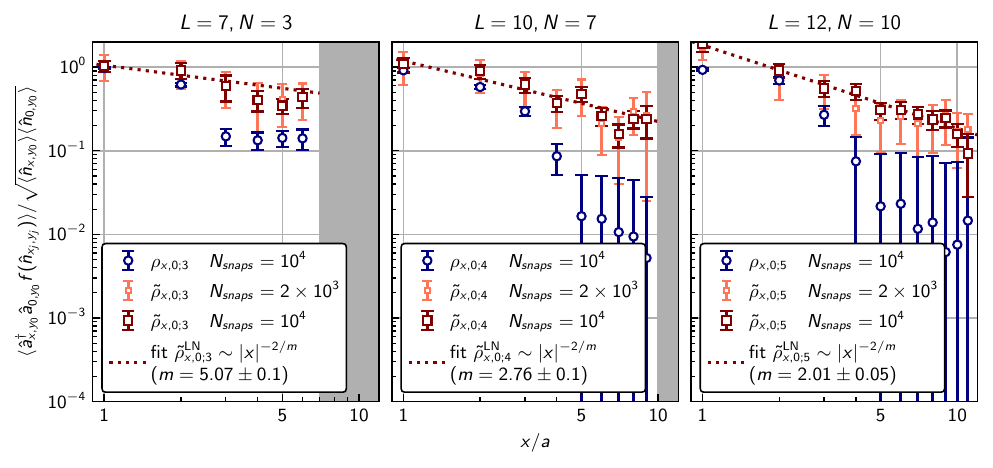}
	\caption{
		Decay of correlations for the original bosons ( $|\rho_{x,x^{\prime};y}| \propto \left\vert \left\langle\hat{a}_{x,y}^{\dagger}\hat{a}^{\nodagger}_{x^{\prime},y} \right\rangle \right\vert$, circles) and the composite bosons (\mbox{$|\tilde{\rho}_{x,x^{\prime};y}| \propto \left\vert \left\langle\hat{b}_{x,y}^{\dagger} \hat{b}^{\nodagger}_{x^{\prime},y}\right\rangle \right \vert$}, squares and triangles) for the system sizes and particle numbers of Fig.~\ref{Fig:Density} with reference at the edge, $x^{\prime}=0$.
		That means, for $\rho$, $f(\hat{n}_{x_j,y_j})\equiv 1$ while for $\tilde{\rho}$, $f(\hat{n}_{x_j,y_j})$ is to be identified with the phase contribution due to the flux attachment in Eq.~\eqref{Eq:LatticeHODLRO}.  
		Already relatively few snapshots ($N_{\rm snaps} = 2\times 10^3$, orange triangles) are sufficient to distinguish the algebraic decay of the correlations $\tilde{\rho}_{x,x^{\prime};y}$ of the composite bosons from the exponential decay of the ordinary correlations $\rho_{x,x^{\prime};y}$.
		Furthermore, for sufficiently large systems and $N_{\rm snaps} = \times 10^4$ (red squares)  a fit of the algebraic decay gives an exponent $-\nicefrac{2}{m}$ consistent with $m=2$.
		The fit is obtained using only intermediate-range points within the bulk of the system.
		The errorbars for the sampling of both the bosonic (blue) and the composite bosonic (orange, red) two-point correlations correspond again to a single standard deviation $\pm \nicefrac{\sigma_{\rho_{x,0;y}}}{\sqrt{N}}$ and $\pm \nicefrac{\sigma_{\tilde{\rho}_{x,0;y}}}{\sqrt{N}}$ respectively.
	}
	\label{Fig:AlgebraicDecay}
\end{figure*}

In our numerical analysis we evaluate the expectation value for the ground state wave function, both using an exact MPS contraction, and analyzing Fock basis snapshots taken by the projective sampling method introduced below in Section \ref{sec:methods}.
\par
To study long-range correlations we fix the reference site to be on the edge of the system, i.e. $x^{\prime}=0$.
Furthermore, we additionally fix $y = y^{\prime} = y_0$ close to the center of the system ($y_0\approx \nicefrac{L}{2}$) to ensure that we eventually probe the bulk of the system. 
\par
The quasi-exact MPS expectation value for the one-particle reduced density matrix $\rho_{x,0; y_0}$ of the ground state reproduces the continuum prediction of Eq. \eqref{exp_continuum} for the system sizes $L=10$ and $L=12$ up to intermediate-range correlations, see Fig.~\ref{Fig:ExponentialDecay} (green diamonds).
As $\ell_B\approx a$, we expect only short-range correlations to be affected by the competition of the characteristic length scales set by the magnetic and the lattice length.
For long distances of the order $|x-x^{\prime}|>\nicefrac{L}{2}$, we observe evident signatures of finite-size effects weakening the decay. 
We believe the presence of an edge mode to be the reason for the deviation from the continuum prediction in this limit.
In support of this, a probe of the correlator along the systems edge (see Appendix \ref{app:B}) yields conclusive evidence for an algebraic decay characteristic for such an edge mode.
For the smallest system, $L=7$, finite size effects dominate and no clear signature of an exponential decay can be found.
\par
We additionally analyze $\rho_{x,0; y_0}$  using the projective two-site sampling from MPS.
We find that already $N_{\rm snaps}=2\times 10^3$ snapshots are sufficient to qualitatively reproduce the exponential decay of $\rho_{x,0; y_0}$ on intermediate scales obtained via the MPS contraction, see Fig.~\ref{Fig:ExponentialDecay} (light-blue circles with errorbars).
The plotted values are the means of the sampling given with an errorbar corresponding to a single standard deviation $\pm \nicefrac{\sigma_{\rho_{x,0;y}}}{\sqrt{N}}$.
In Appendix \ref{app:A} we justify this choice and analyze the convergence behavior of the sampling to the expectation value in the limit $N_{\rm snaps}\rightarrow \infty$ and evaluate its statistical accuracy.
Using the data points of the sampling, we perform an exponential fit for all system sizes yielding a correlation length $\xi \approx \nicefrac{3a}{2}$ which is slightly larger than the magnetic length $\ell_B \approx a$.

\subsection{Emergence of HODLRO}
The central physical question addressed in this paper is the existence of HODLRO in fractional Chern insulators.
As introduced in Section~\ref{sec:proposal}, HODLRO can be understood as the algebraic long-range order of emergent objects.
To achieve this, the original bosons are transformed to \textit{composite bosons} via a singular gauge transformation~\cite{Wilczek1982,Arovas1984,Jain1989}, which corresponds to the attachment of flux quanta to the fundamental bosons.
In the case of the Laughlin state at $\nu = \nicefrac{1}{m}$, each particle is endowed with exactly $m$ flux quanta.
The attached flux quanta cause the composite object to acquire an additional statistical angle of $m\pi$. 
Hence, independent of the filling and the underlying particles the composite particles obtained by this procedure are always bosonic.
\par
For varying $x,x^{\prime}$ and fixed $y=y^{\prime}=y_0$ the composite boson correlations in the continuum Laughlin state follow a power-law scaling of the form
\begin{equation}
	\tilde{\rho}_{x,x^{\prime}; y_0}^{\rm LN} \sim |x-x^{\prime}|^{-2\nu} = |x-x^{\prime}|^{-\nicefrac{2}{m}},
	\label{Eq:AlgebraicDecay}
\end{equation}
indicating HODLRO in the Laughlin state~\cite{Girvin1988,Read1989}.	
In contrast to the ordinary bosonic correlation, the transformed composite bosonic expression cannot readily be evaluated using direct contractions of the MPS representation of the ground state, because one explicitly needs to know the occupation numbers $n_{\boldsymbol{k}}$ on all lattice points which do not coincide with $\boldsymbol{j}=(x,y_0)$ and $\boldsymbol{l}=(x^{\prime},y_0)$, while simultaneously determining the bosonic one-particle coherence on the two remaining sites.
However, the projective two-site sampling algorithm we introduce in this paper grants access to exactly this information.
Applying this algorithm, we can generate snapshots to evaluate the expectation value with in principle arbitrary accuracy.
Moreover, this scheme precisely emulates the measurement protocol for a quantum gas microscopy experiment as proposed in Section~\ref{sec:proposal} above.
%

Analogously to the analysis of the bare correlator, we compute the normalized transformed, multi-particle two-point correlation function
\begin{equation}
	\tilde{\rho}_{x,x^{\prime}; y} =\left\langle \hat{b}_{x,y}^{\dagger} \hat{b}^{\nodagger}_{x^{\prime},y}\right\rangle/\sqrt{\left\langle \hat{n}_{x,y}\right\rangle \left\langle \hat{n}_{x^{\prime},y}\right\rangle},
\end{equation}
for all system sizes with the reference site fixed at the center of one edge, $x^{\prime}=0, y_0\approx \nicefrac{L}{2}$.
Already for $N_{\rm snaps}=2\times 10^3$ snapshots we find a clear difference between the exponential decay for the bare correlator $\rho$ and the algebraic decay of the transformed expression $\tilde{\rho}$ for all system sizes, see Fig.~\ref{Fig:AlgebraicDecay}.
This clearly indicates the presence of HODLRO in the ground state of our lattice model for a number of snapshots realistically achievable in cold atom experiments.
Moreover, the emergence of the power-law scaling is robust to the choice of the reference site $x^{\prime}$, which is shown in more detail in Appendix~\ref{app:C}.
\par
Building on those qualitative observations, we perform a fit of the analytically predicted decay in Eq.~\eqref{Eq:AlgebraicDecay} to our sampled data and find an excellent agreement of the exponent with the analytically predicted value for $N_{\rm snaps}=10^4$ snapshots in the largest systems, $L=12$.
We find a power-law scaling with an exponent $m_{L=12}=2.01\pm 0.05$ which agrees perfectly with the predicted value $m=2$.
For the smaller systems, $L= 7, 10$, we find that we can still distinguish algebraic and exponential decay, however the fit to the analytical prediction becomes less accurate.
\par
In the case of the smallest system studied here ($L=7$), the situation gets even more complicated as an obvious bulk region is hardly identifiable and edge effects are believed to significantly influence the results.
Interestingly, while the exponential decay is destroyed by finite size effects, the algebraic decay in the composite bosonic correlations is robust against them, signaling an underlying universal property of the state responsible for its emergence.
We conclude that such experiments constitute a prime platform giving insights about the nature of the intrinsic topological order of the state by pure analysis of the ground state.

\section{Methods\label{sec:methods}}
We now turn to the discussion of the sampling protocol to simultaneously generate one- and two-site snapshots from a MPS which was instrumental for our preceding numerical analysis of HODLRO.
First, we provide the reader with a brief review of some features of MPS which are essential to our snapshot set-up and the idea to numerically emulate experimental cold atom quantum gas microscope measurements. 
Afterwards, as the single-site/two-site algorithm is a straightforward generalization of the single-site sampling, we continue by shortly reviewing the idea of the single-site method.
Building up on this, in the last subsection, we discuss the full single-site/two-site variant of the algorithm.

\subsection{Matrix Product State Set-up}
We consider a tensor product Hilbert space $\mathcal{H}$ of $L$ sites with local basis states $\{{\ket{\sigma_{j,1}},\ket{\sigma_{j,2}},...,\ket{\sigma_{j,p}}}\}$, where $p$ denotes the local physical dimension, so that the full Hilbert space is given by
\begin{equation}
    \begin{aligned}
        \mathcal{H} = \bigotimes^L_{j=1} \mathcal{H}_j, \qquad \mathcal{H}_j = \operatorname{span}(\{\ket{\sigma_j}\}).
    \end{aligned}
\end{equation}
The wavefunction of an arbitrary pure state $\ket{\psi}\in \mathcal{H}$ can be represented by a product of matrices such that
\begin{equation} \label{eq:MPS}
        \ket{\psi}=\sum_{\substack{\sigma_1,...,\sigma_L \\ \mu_0,...,\mu_L}} M_{\mu_0,\mu_1}^{\sigma_1} \cdot \cdot \cdot M_{\mu_{L-1},\mu_L}^{\sigma_L} \ket{\sigma_1,...,\sigma_L},
\end{equation}
which yields the MPS representation of the state.
Here, the site-local matrix $M_j^{\sigma_j}$ is unique only up to an invertible linear map.  We can exploit this gauge freedom to rewrite the site tensor in its \textit{left-canonical}/\textit{right-canonical}  form $A_j^{\sigma_j}$/$B_j^{\sigma_j}$, which is defined by
\begin{equation}
    \begin{aligned}
           \sum_{\sigma_j} (A_j^{\sigma_j})^{\dagger}A_j^{\sigma_j} &= \mathbb{1}_j, \\
            \sum_{\sigma_j}(B_j^{\sigma_j})^{\dagger}B_j^{\sigma_j} &= \mathbb{1}_j.
    \end{aligned}
\end{equation}
Left-/right-canonical MPS are defined by requiring that they consist of left-/right-normalized tensors only.
We can also construct a mixed-canonical MPS by fixing a site $j$ for which the tensor $M_j^{\sigma_j}$ remains unchanged while we demand all site tensors to the left/right to be left-/right-normalized tensors.
\begin{figure}[b]
	\centering
	\includegraphics{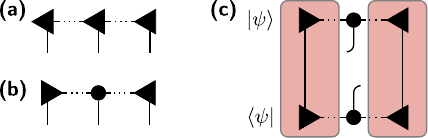}
	\caption{\textbf{(a)} Schematic sketch of the tensor-network of the right-canonical MPS representation of an arbitrary state $\ket{\psi}\in\mathcal{H}$. \textbf{(b)} Now the state is shown in its mixed-canonical form where all sites to the left/ right of the active site are in the left-/ right-canonical gauge (blue left-/ right-orientated triangle nodes).
		\textbf{(c)} The one-site reduced density matrix $\hat{\rho}_j$ in its tensor network graph representation. Due to the mixed-canonical form of the state the tensor contraction left/ right to the active site (round blue node) will give the identity (light-orange boxes). $\hat{\rho}_j$ gives access to the full site-local probability distribution.}
	\label{Fig:CanonicalForm}
\end{figure}
We then call site $j$ the \textit{active site}.  
The canonical form of the MPS tensor-network is visualized in Fig.~\ref{Fig:CanonicalForm}.
The canonical form is especially advantageous if we want to compute the expectation value of an arbitrary site-local operator $\hat{o}_j$
\begin{equation}
	\begin{aligned}
			\langle \hat{o}_j \rangle = \text{Tr}_j \{ \hat{\rho}_j \hat{o}_j\} = \sum_{o_j} \langle o_j | \hat{\rho}_j \hat{o}_j|o_j \rangle = \sum_{o_j} \rho^{o_j o_j} o_j,
	\end{aligned}
\end{equation}
as the one-site reduced density matrix $\hat{\rho}_j$ becomes trivial and grants access to the full probability distribution for the eigenvalue spectrum of the operator given by the diagonal elements $\rho^{o_j o_j}$. For a detailed discussion of MPS and its technicalities we remit to \cite{Schollwock2011}.
\begin{figure*}[ht]
	\centering
	\includegraphics{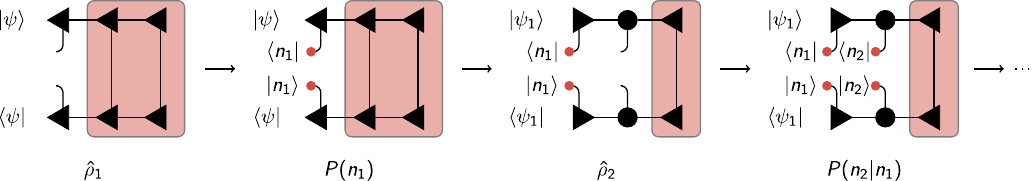}
	\caption{Schematic sketch of the one-site perfect sampling of site-local density operators $\hat{n}_j$ for the MPS representation of a state $\ket{\psi}$ on a lattice with 3 sites.  
		Starting at site 1 in the right-canonical gauge of the MPS we obtain the probability distribution $\sum_{n_1}P(n_1)=1$ via the diagonal elements of the one-site reduced density matrix $\rho^{n_1 n_1} =\braket{n_1|\hat{\rho}_1|n_1}=P(n_1)$. 
		Upon drawing a certain eigenvalue $n_1$ from the probability distribution the state is projected in the corresponding site-local product state.
		By moving the active site of the projected state to the second site one can read out the conditioned probability distribution $\sum_{n_2}P(n_2|n_1)=1$ via $\rho^{n_2 n_2}=\braket{n_1,n_2|\hat{\rho}_2|n_1,n_2}$. 
		Finally, one obtains a single-full lattice snapshot of the form $(n_1,n_2,n_3)$ from $P(n_1,n_2,n_3)=P(n_1)\cdot P(n_2|n_1)\cdot P(n_3|n_2,n_1)$.}
	\label{Fig:One-Site-Alg}
\end{figure*}
\subsection{Quantum Simulator Set-up}
As we are interested in Fock basis snapshots, from now on we work in the occupation number basis and choose $\hat{o}_j = \hat{n}_j$, but in principle the following scheme works for arbitrary operators.
\par
Taking Fock basis snapshots of the pure state $\ket{\psi}$ corresponds to simultaneously measuring the site-local occupation number operator $\hat{n}_j$ on each lattice site.
We can decompose each $\hat{n}_j$ using local projection operators $\hat{\mathcal{P}}_{j}^{n_j}$ onto the eigenspaces of the corresponding measurement outcomes $n_{j}$, where $n_j \in [0, 1, \hdots, p-1]$
\begin{equation}
	\hat{n}_j = \sum_{n_j=0}^{p-1} n_{j} \hat{\mathcal{P}}_{j}^{n_j}.
\end{equation}
The global particle number operator $\hat{N} = \sum_{i=1}^{L} \hat{n}_j$ can be rewritten as
\begin{equation}
		\hat{N} = \sum_{j=0}^{L}\sum_{n_j=0}^{p-1}n_j\hat{\mathcal{P}}_j^{n_j} ,
\end{equation}
and consequently, the measurement outcome of a single snapshot is given by a tuple $\left(n_{1}, \hdots,n_{L}\right)$, which accordingly corresponds to a pattern of projectors of the initial state into site-local eigenspaces of the density operator. 
\subsection{Sampling Single-Site Operators}
Given the MPS representation of an arbitrary quantum state, we can emulate state-of-the-art quantum simulator measurements by drawing independent snapshots employing the perfect sampling scheme first introduced by Ferris and Vidal~\cite{Ferris2012}.
Contrary to other sampling procedures such as Markov chain Monte Carlo sampling, this algorithm produces perfectly uncorrelated samples (and hence the name perfect sampling). This is computationally advantageous, as we do not need to account for additional equilibration and autocorrelation times. 
The perfect sampling scheme for a global lattice operator, which is decomposable in site-local observables is a straightforward successive application of the canonical form and tensor contractions which has been discussed in the framework of MPS in~\cite{Buser2022}.
\par
For a treatment of the ground state of the Hofstadter-Bose-Hubbard model, we also need to account for the $\mathrm{U}(1)$-symmetry of the system associated with the conservation of the particle number $\hat{N}$ constraining tensor manipulations.
To clarify this, we briefly review the key steps of the single-site algorithm here.
\par
The idea of projectively sampling is best understood, if we first consider the expectation value of the global observable $\Hat{N}$ on our lattice Hilbert space $\mathcal{H}$, which we can express as
\begin{equation} \label{eq:expectation value}
    \begin{aligned}
        \braket{\psi|\Hat{N}|\psi} = \sum_{\vec{n}\in \mathcal{N}} \braket{\psi|\vec{n}}\braket{\vec{n}|\Hat{N}|\psi}.
    \end{aligned}
\end{equation}
Here $\mathcal{N}$ denotes the set of all $p^L$ possible outcome configuration of the outcome tuple $\vec{n} \equiv (n_{1},...,n_{L})$ and $\ket{\vec{n}}\equiv \ket{n_1}\otimes \ket{n_2}\otimes \cdot \cdot \cdot \otimes \ket{n_L}$ is a product state with $n_{i} = 0,1,...,p-1$ possible instances labeling the elements of a local orthonormal basis $\{\ket{n_{i}}\}$. 
This expression can be interpreted in the following way.
\begin{figure*}[ht]
	\centering
	\includegraphics{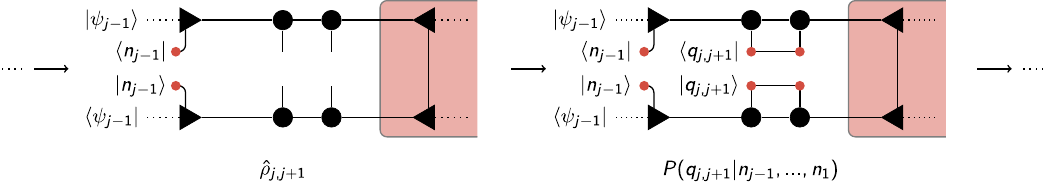}
	\caption{Schematic sketch of the modification of the one-site algorithm towards the mixed one-site/ two-site sampling to evaluate expectation values of the form $\left\langle \hat{n}_1...\hat{n}_{j-1}\hat{q}_{j,j+1}\hat{n}_{j+2}...\hat{n}_L \right\rangle$.
		After $j-1$ single-site steps we bring the projected MPS state $\ket{\psi_{j-1}}$ in a canonical gauge where both site $j$ and site $j+1$ are the active sites (round blue nodes). 
		We are now interested in the probability distribution $\sum_{q_{j,j+1}} P(q_{j,j+1}|n_{j-1},...,n_1)=1$ which is now accessible via the diagonal elements of the two-site reduced density matrix $\rho_{j,j+1}^{q_{j,j+1} q_{j,j+1}}=\braket{n_1,...,n_{j-1},q_{j,j+1}|\hat{\rho}_{j,j+1}|n_1,...,n_{j-1},q_{j,j+1}}=P(q_{j,j+1}|n_{j-1},...,n_1)$.
		After drawing $q_{j,j+1}$ in accordance to $P(q_{j,j+1}|n_{j-1},...,n_1)$ the scheme is continued by performing single-site measurements starting at site $j+2$ until we obtain a full-lattice snapshot of the type $(n_1,...,n_{j-1},q_{j,j+1},n_{j+2},...,n_L)$.
	}
	\label{Fig:Two-Site-Sampling}
\end{figure*}
We introduce the probability $\vert \braket{\vec{n}|\psi}\vert ^2 \equiv P(\vec{n})$
of projecting $\ket{\psi}$ into the product state $\ket{\vec{n}}$, and the \textit{estimator} $\braket{\vec{n}|\Hat{N}|\psi}/\braket{\vec{n}|\psi} \equiv E(\vec{n})$ such that:
\begin{equation}
    \begin{aligned}
        \braket{\psi|\Hat{N}|\psi} = \sum_{\vec{n}\in \mathcal{N}} P(\vec{n}) E(\vec{n}).
    \end{aligned}
\end{equation}
Evidently, $\sum_{\vec{n}\in \mathcal{N}}P(\vec{n})=1$ and $|\mathcal{N}|=p^L$.
In order to approximate $\braket{\psi|\Hat{N}|\psi}$, consider only a subset of configurations $\tilde{\mathcal{N}}\subset \mathcal{N}$ with $\vert \Tilde{\mathcal{N}} \vert <p^L$. 
Then upon normalizing with $Z= \sum_{\vec{n} \in \tilde{\mathcal{N}}}P(\boldsymbol{n})$ we can use the following approximation of the exact expectation value
\begin{equation} \label{SamplingExp}
    \begin{aligned}
        \braket{\psi|\Hat{N}|\psi} \approx \frac{1}{Z} \sum_{\vec{n}\in \tilde{\mathcal{N}}}P(\vec{n})E(\vec{n}).
    \end{aligned}
\end{equation}
If we now suppose that the $|\tilde{\mathcal{N}}|$ configurations of the subset $\tilde{\mathcal{N}}$ were drawn randomly from all possible configurations $|\mathcal{N}|$ in accordance to the probability distribution $P(\vec{n})$, we can rewrite this approximation as
\begin{equation}
    \begin{aligned}
        \frac{1}{Z}\sum_{\vec{n}\in \tilde{\mathcal{N}}}P(\vec{n})E(\vec{n}) = \frac{1}{|\tilde{\mathcal{N}}|}\sum_{\vec{n}\in\tilde{\mathcal{N}}}E(\vec{n}).
    \end{aligned}
\end{equation}
That is, we estimated $\braket{\psi|\Hat{N}|\psi}$ by means of $|\tilde{\mathcal{N}}|$ independent samples from the random variable $(P(\vec{n}),E(\vec{n}))$.
Instead of performing the full contraction in order to obtain Eq.~\eqref{eq:expectation value}, we can sample the observable $\Hat{N}$ by drawing a finite number of samples from the probability distribution.
Based on this result, constructing the probability distribution $P$ is the key to the projective sampling algorithm.
As elicited above, MPS are an ideal starting point to construct $P$, due to the isometric character of the site tensors which enable us to express the MPS in the canonical/unitary form giving access to the full probabilities through the one-site density matrices. 
\par
Hence, the first step of the algorithm is bringing $\ket{\psi}$ in the right-canonical form
\begin{equation}
    \begin{aligned}
        \ket{\psi} = \sum_{n_1}\sum_{n_2}\cdot \cdot \cdot \sum_{n_L} B^{n_1}_1 B^{n_2}_2 \cdot \cdot \cdot B^{n_L}_L \ket{\vec{n}}.
    \end{aligned}
\end{equation}
Subsequently, we select the first site-local density operator $\Hat{n}_1$ and break it down in its $p$ projector constituents
\begin{equation}
    \begin{aligned}
        \Hat{n}_1 = \sum_{n_1}n_{1}\hat{\mathcal{P}}_{1}^{n_1}.
    \end{aligned}
\end{equation}
On the tensor-network level, this can be achieved following Ref.~\cite{Singh_2011} by manipulating the local rank 4 site tensor of the density operator in accordance with the constraints given by particle number conservation until we have a compactified rank 2 tensor.
The rank 2 tensor is a Hermitian matrix and thus, we can apply an eigenspace decomposition. 
\par
We obtain the probability $P(n_{1})$ to draw $n_{1}$ via 
\begin{equation}
    \begin{aligned}
        P(n_{1}) \equiv \langle n_1 | \hat{\rho}_1 | n_1 \rangle,
    \end{aligned}
\end{equation}
where $\ket{n_1} \equiv \hat{\mathcal{P}}_{1}^{n_1}\ket{\psi}$.
Doing this for all $p$ possible measurement outcomes yields the full probability distribution satisfying
\begin{equation}
    \begin{aligned}
        \sum_{n_1} P(n_{1}) = 1.
    \end{aligned}
\end{equation}
We randomly choose one of the $p$ eigenvalues representing the measurement outcome in accordance to their probability distribution $P(n_1)$.
\par
Finally, we project the MPS site-tensor in the drawn site-local product-state and normalize it by means of a singular value decomposition in order to shift the canonical center towards the second site
\begin{equation}
    \begin{aligned}
        \ket{\psi_1} = \frac{\Hat{\mathcal{P}}_{1}^{n_1}\ket{\psi}}{\sqrt{\braket{\psi|\Hat{\mathcal{P}}_{1}^{n_1}|\psi}}}.
    \end{aligned}
\end{equation}
%
%
This one-site sampling step is now repeated for each of the $L$ lattice sites. 
The initial state $\ket{\psi_k}$ of the $k$th sampling step is given by
\begin{equation}
	\begin{aligned}
		\ket{\psi_k} = \frac{\hat{\mathcal{P}}_{k}^{n_k}\ket{\psi_{k-1}}}{\sqrt{\braket{\psi|\Hat{\mathcal{P}}_{k}^{n_k}|\psi_{k-1}}}}.
	\end{aligned}
\end{equation}
The sampling probabilities at each step are conditioned by all already projected site tensors and therefore the final probability distribution $P(n_1,...,n_L) \equiv P(\vec{n})$ is given by
\begin{equation}
    \begin{aligned}
       P(\vec{n}) = P(n_{1})\cdot P(n_{2}|n_{1}) \cdot \cdot \cdot P(n_{L}|n_{{L-1}},...,n_{2},n_{1}).
    \end{aligned}
\end{equation}
Consequently, by sweeping over the full lattice we obtain a single snapshot-tuple $(n_{1},...,n_{L})$ drawn according to $P(\vec{n})$. 
The whole perfect sampling scheme for MPS is illustrated in Fig. \ref{Fig:One-Site-Alg}.
\subsection{Sampling Two-Site Operators}
In the context of revealing HODLRO we are interested in sampling an operator expectation value of the type
\begin{equation}\label{Eq:One-Site/Two-Site Operator}
	\left\langle \hat{a}_{\boldsymbol{j}}^{\dagger} \hat{a}_{\boldsymbol{l}}^{\nodagger} \prod_{\boldsymbol{k}\neq \boldsymbol{j},\boldsymbol{l}} \hat{n}_{\boldsymbol{k}} \right\rangle
\end{equation}
where we adopted the notation of Sec.~\ref{sec:proposal}, i.e. a bold letter corresponds to a discrete coordinate tuple $(x,y)$. 
Eq.~\eqref{Eq:One-Site/Two-Site Operator} underlines that we want to simultaneously sample Fock basis snapshots on all but two sites, and on the two remaining sites we are interested in the bosonic one-particle correlation function.
Such operators are not accessible through the introduced one-site perfect sampling scheme, however, by generalizing the method in fact any kind of multi-site observable can be probed.
\par
Here we will present the modification needed in order to sample operators of the type of Eq. \eqref{Eq:One-Site/Two-Site Operator}, i.e. we need to additionally extract probabilities for eigenvalues of non-local two-site operators.
Analogously to the single-site case, these probabilities are contained within the two-site reduced density matrix 
\begin{equation}\label{Eq:Two-Site RDM}
	\begin{aligned} 
		\hat{\rho}_{\boldsymbol{j},\boldsymbol{l}} = \text{Tr}_{{\boldsymbol{k}} \neq \boldsymbol{j},\boldsymbol{l}} \hat{\rho}.
	\end{aligned}
\end{equation}
\par
We can simplify sampling Eq. \eqref{Eq:Two-Site RDM}  by exploiting the invariance of the trace under permutations $\Hat{P}$, which are involutory, i.e. fulfill $\Hat{P}\Hat{P}=\mathbb{1}$.
Hence, in order to sample an arbitrary non-local two-site observable $\hat{q}_{\boldsymbol{j},\boldsymbol{l}}$ with $|\boldsymbol{j}-\boldsymbol{l}|>1$, we can use the permutation $\hat{P}$ to change the ordering of the sites  $\{\boldsymbol{j},\boldsymbol{l}\} \rightarrow \{\boldsymbol{j^{\prime}},\boldsymbol{l^{\prime}}\}$, such that $|\boldsymbol{j}^{\prime}-\boldsymbol{l}^{\prime}|=1$. 
From a MPS perspective this is beneficial, as we do not need to account for non-local operations. 
That means, we can safely assume that we are interested in sampling the following set of local operators $(\hat{n}_{\boldsymbol{1}},...,\hat{n}_{\boldsymbol{j}-1},\hat{q}_{\boldsymbol{j},\boldsymbol{j}+1},\hat{n}_{\boldsymbol{j}+2},...,\hat{n}_{\boldsymbol{L}})$
where the coordinate arithmetic $\boldsymbol{j}\pm 1$ corresponds to the two adjacent lattice points of site $\boldsymbol{j}$ in an arbitrary mapping of the two-dimensional lattice to a one-dimensional chain.
In this way, up until to the $j$th sampling step we can follow the single-site procedure, i.e. we begin by sampling a pattern of site-local occupation number eigenvalues of the form $(n_{\boldsymbol{1}},...,n_{\boldsymbol{j}-1})$ in accordance to the probability distribution $P(n_{\boldsymbol{1}},...,n_{\boldsymbol{j}-1})$.
\par
Next, we take the projected state $\ket{\psi_{\boldsymbol{j}-1}}$ and bring it in a mixed canonical form, but this time both site $\boldsymbol{j}$ and $\boldsymbol{j}+1$ form the canonical center.  
This enables us to trivially compute the two-site reduced density matrix granting access to the conditioned sampling probabilities.
\par
The next step is to spectrally decompose $\Hat{q}_{\boldsymbol{j},\boldsymbol{j}+1}$.
On the tensor network level $\Hat{q}_{\boldsymbol{j},\boldsymbol{j}+1}$ is now a rank 6 tensor which increases the complexity in fusing it down to its compactified rank 2 form yielding
\begin{equation}
    \begin{aligned}
        \Hat{q}_{\boldsymbol{j},\boldsymbol{j}+1} = \sum_{q_{\boldsymbol{j},\boldsymbol{j}+1}}q_{\boldsymbol{j},\boldsymbol{j}+1}\hat{\mathcal{P}}_{\boldsymbol{j},\boldsymbol{j}+1}^{q_{\boldsymbol{j},\boldsymbol{j}+1}}.
    \end{aligned}
\end{equation}
Analogously to the single-site step, we obtain the probability $P(q_{\boldsymbol{j},\boldsymbol{j}+1}|n_{\boldsymbol{j}},...,n_{\boldsymbol{L}})\equiv \bar{P}$ to draw the eigenvalue $q_{\boldsymbol{j},\boldsymbol{j}+1}$ via 
\begin{equation}
	\begin{aligned}
		\bar{P} = \langle n_{\boldsymbol{1}},...,n_{\boldsymbol{L}},q_{\boldsymbol{j},\boldsymbol{j}+1}| \hat{\rho}_{\boldsymbol{j},\boldsymbol{j}+1} | n_{\boldsymbol{1}},...,n_{\boldsymbol{L}},q_{\boldsymbol{j},\boldsymbol{j}+1} \rangle,
	\end{aligned}
\end{equation}
where $\ket{n_{\boldsymbol{1}},...,n_{\boldsymbol{L}},q_{\boldsymbol{j},\boldsymbol{j}+1}} \equiv \hat{\mathcal{P}}_{\boldsymbol{j},\boldsymbol{j}+1}^{q_{\boldsymbol{j},{\boldsymbol{j}}+1}} \ket{n_{\boldsymbol{1}},...,n_{\boldsymbol{L}}} $.
\par
After having done that for all possible measurement outcomes of the two-site operator, once again we obtain an eigenvalue $q_{\boldsymbol{j},\boldsymbol{j}+1}$ by randomly drawing from $\bar{P}$ and the updated snapshot takes the form $(n_{\boldsymbol{1}},...,n_{\boldsymbol{j}-1},q_{\boldsymbol{j},\boldsymbol{j}+1})$.

The two-site step is completed by the projection on the corresponding two-site product state:
\begin{equation}
    \begin{aligned}
        \ket{\psi_{\boldsymbol{j}}}=\Hat{\mathcal{P}}_{\boldsymbol{j},\boldsymbol{j}+1}^{q_{\boldsymbol{j},\boldsymbol{j}+1}}\ket{\psi_{\boldsymbol{j}-1}},
    \end{aligned}
\end{equation}
and ultimately, by moving the canonical center towards the site $\boldsymbol{j}+2$. 
The extended perfect sampling algorithm in its tensor network notation is shown in Fig.~\ref{Fig:Two-Site-Sampling}.
\par
From here onward, the procedure is just the plain one-site algorithm again until we finally obtain $(n_{\boldsymbol{1}},...,n_{\boldsymbol{j}-1},q_{\boldsymbol{j},\boldsymbol{j}+1},n_{\boldsymbol{j}+2},...,n_{\boldsymbol{L}})$.
%
How a snapshot of this form can be used to probe HODLRO is discussed in detail in Appendix \ref{app:E} and \ref{app:F}.

\section{Conclusion and Outlook\label{sec:conclusion}}
The main result of our work is the demonstration that state-of-the-art quantum simulation platforms, and cold atoms in particular, can be used to directly detect HODLRO in fractional Chern insulators. 
This finding is based on strong numerical evidence for the existence of HODLRO in a lattice analog of the $\nu=\nicefrac{1}{2}$-Laughlin state, which has far-reaching consequences for topologically ordered lattice systems.
We explicitly generalized existing continuum results to lattice systems and described how correlations of the emergent composite bosons give rise to HODLRO in this context.
Such correlators are of interest to probe the condensation of emergent composite bosons and to deepen our understanding of the intrinsic topological order in the quantum state.
\par
In view of the current abilities of cold atom quantum simulators, we proposed an experimentally accessible scheme to extract long-range correlations of the form $\braket{\hat{a}_{\boldsymbol{j}}^{\dagger} \hat{a}_{\boldsymbol{l}}^{\nodagger} \prod_{\boldsymbol{k}\neq \boldsymbol{j},\boldsymbol{l}} \hat{n}_{\boldsymbol{k}}}$.
Our proposal is based on realizing a bilayer system which is coupled along one edge to realize an effective extended single-layer Hofstadter-Bose-Hubbard model.
This allows to measure non\hyp local coherences $\sim\hat{a}^{\dagger}_{\boldsymbol{j}}\hat{a}^{\nodagger}_{\boldsymbol{l}}$ in current setups.
\par
Our numerical results are based on the extension of the existing perfect sampling scheme~\cite{Ferris2012} to allow to projectively sample more complex, multi-particle correlation functions like HODLRO.
The application of this sampling algorithm allowed us to probe HODLRO, thus demonstrating that this concept persists in lattice systems accessible to near-term quantum simulators.
In particular, we showed that already relatively few snapshots ($N_{\rm{snaps}}=\mathcal{O}(10^3)$) are sufficient to distinguish the exponentially decaying correlations of the underlying bosons from the quasi-long-ranged power-law decay of the composite boson correlations.
Furthermore, we were able to resolve the exponent of the algebraic decay which is directly related to the filling factor of the Laughlin state and gives direct qualitative insights about the intrinsic topological order of the state, specifically its K-matrix.

The recent progress in studying FQH states in cold atom experiments~\cite{Leonard2023} calls for additional ways to probe these states directly.
The snapshot-based protocol discussed here provides a way to explore the unconventional correlators needed, for example, to reveal the intrinsic topological order of such states.
Having revealed the exotic correlations present in such systems, a next step would be further investigations of their origin.
Microscopically, computing the entire one-particle reduced density matrix for the composite bosons employing the snapshot protocol could give further insights into the speculated condensation of the composite bosons in the Laughlin state.
Another intriguing possibility is to explore the fate of HODLRO in systems out-of equilibrium.
While we restricted our analysis to a particularly simple FQH state, we believe that the concepts introduced here are also applicable to more exotic FQH states, like for example the Pfaffian state~\cite{Moore1991} or parafermion states~\cite{Read1999}.
Furthermore, applying similar ideas to other systems exhibiting topological order, like chiral spin liquids~\cite{Kalmeyer1987}, might deepen our understanding of such exotic states of matter.
These more complex states should be in principle accessible to quantum simulator platforms like ultracold atoms or superconducting qubits where our approach can be readily applied.

\begin{acknowledgments}
	\ 
	The authors would like to thank Nathan Goldman, Julian L\'eonard, and Eugene Demler for fruitful discussions.
	We acknowledge support by the Deutsche Forschungsgemeinschaft (DFG, German Research Foundation) under Germany’s Excellence Strategy-426 EXC-2111-390814868.
	FAP and FG acknowledge funding by the Deutsche Forschungsgemeinschaft (DFG, German Research Foundation) via Research Unit FOR 2414 under project number 277974659 and from the European Research Council (ERC) under the European Union’s Horizon 2020 research and innovation programm (Grant Agreement no 948141) -- ERC Starting Grant SimUcQuam.
\end{acknowledgments}

\appendix
\section{Singular Gauge Transformation} \label{App:SGT}
Consider a single charged particle in a plane with position $\boldsymbol{r^{\prime}}=(0,0)$.
We can endow this particle with a magnetic flux quantum $\Phi_0$ by introducing a differential form $\mathcal{A}$ which satisfies
\begin{equation}
	\boldsymbol{\nabla} \wedge \mathcal{A} = \mathcal{B}(\boldsymbol{r}) = \Phi_0 \delta^{(2)}(\boldsymbol{r}).
\end{equation} 
where $\wedge$ is the exterior product in two spatial dimensions, i.e. $\boldsymbol{\nabla} \wedge \mathcal{A}=\partial_x \mathcal{A}_y - \partial_y \mathcal{A}_x$.
$\mathcal{A}$ generates an infinitesimal, singular magnetic field which vanishes everywhere but on the particle itself.
We find $\mathcal{A}$ by solving 
\begin{equation}
	\int_{\mathbb{R}} dx\int_{\mathbb{R}} dy(\partial_x\mathcal{A}_y - \partial_y \mathcal{A}_x)=1.
\end{equation}
Imposing symmetry and using \mbox{$\int_{\mathbb{R}}du \frac{1}{1+u^2}=\arctan{(u)}|^{+\infty}_{-\infty}=\pi$} we find
\begin{equation} \label{Eq:1SGT}
	\begin{aligned}
	\mathcal{A} &= \frac{\Phi_0}{2\pi} \left( \frac{-y}{x^2 + y^2}\hat{\boldsymbol{x}} + \frac{x}{x^2 + y^2}\hat{\boldsymbol{y}} \right) \\ &=\frac{\Phi_0}{2\pi}\boldsymbol{\nabla}\arctan{\left(\frac{y}{x}\right)}.
	\end{aligned}
\end{equation}
In order to attach $m$ multiple flux solenoids we can simply write $m\mathcal{A}$.
Now consider a system with $N$ particles parameterized by a set of position vectors $\{\boldsymbol{r}_1,...,\boldsymbol{r}_N\}$. 
We look for a transformation such that each particle acquires $m$ flux tubes relative to all remaining charges which is found as the multi-particle extension of~\eqref{Eq:1SGT}
\begin{equation}
	\mathcal{A}_j(x_j,y_j)=\frac{m}{2\pi}\sum_{k\neq j} \boldsymbol{\nabla}_j \arctan{\left(\frac{y_k-y_j}{x_k-x_j}\right)}
\end{equation}
Using the complex representation of the position vectors, i.e. $z_k=x_k+iy_k$, we can identify
\begin{equation}
	\arctan{\left(\frac{y_k-y_j}{x_k-x_j}\right)}=\arg{(z_k-z_j)}
\end{equation}
which via $\ln{(z)}=\ln{(|z|)}+i\arg{(z)}$ is just $\mathrm{Im}\{\ln{(z_k-z_j)}\}$ and it follows the expression for the singular gauge transformation used in the main text
\begin{equation}
	\mathcal{A}_j(z_j) = \frac{m \Phi_0}{2\pi}\sum_{k\neq j} \boldsymbol{\nabla}_j \mathrm{Im}\{\mathrm{ln}(z_k-z_j)\}.
\end{equation}
If we consider the Laughlin wavefunction \mbox{$\psi_{\rm{LN}}=\prod_{j<k}(z_j-z_k)^m \mathrm{e}^{-\frac{1}{4}\sum_l |z_l|^2}$} such a transformation on each particle brings us to
\begin{equation}
	\psi_{LN} \xrightarrow{\mathcal{A}} \mathrm{e}^{\sum_j 	\int d\boldsymbol{r}_j \mathcal{A}_j(z_j)}\psi_{LN}=\prod_{j<k}|z_j-z_k|^m \mathrm{e}^{-\frac{1}{4}\sum_l |z_l|^2}.
\end{equation}
Independent of $m$, the transformed wave function is purely real and symmetric and hence, describes a bosonic composite of flux quanta and a charged particle.
\section{Gauge Independence} \label{app:D}
We introduced the Hofstadter-Bose-Hubbard model in Landau gauge and hence, all calculations have been carried out using this particular gauge choice. 
It is straightforward to show that the emergence of HODLRO found in Section \ref{sec:Numerical Analysis} is independent of the gauge.
Consider the original two-point correlation function
\begin{equation} \label{eq:D1}
	\rho_{x, x^{\prime}; y, y^{\prime}}= \left\langle \hat{a}^{\dagger}_{x,y}\hat{a}^{\nodagger}_{x^{\prime},y^{\prime}} \right\rangle.
\end{equation}
Under a gauge transformation on the physical vector potential $\boldsymbol{A}$ of the form
\begin{equation}
	A_{x,y}^{k} \longrightarrow A_{x,y}^{k} - \partial^k \chi_{x,y}
\end{equation}
the bosonic annihilation (creation) operators transform as
\begin{equation}
	\hat{a}^{(\dagger)}_{x,y}\longrightarrow e^{(-)i\chi_{x,y}}\hat{a}^{(\dagger)}_{x,y}.
\end{equation}
\\
For instance going from Landau gauge $A_{x,y;L}=2\pi\alpha\left( -y,0\right)$ to the symmetric gauge $A_{x,y;S}=\pi\alpha \left(-y,x\right)$ corresponds to a gauge transformation $\chi_{x,y}=-\pi\alpha xy$.
Consequently, the correlation function of Eq. \ref{eq:D1} simply picks up an additional local phase under general gauge transformations $\chi_{x,y}$ maintaining a homogeneous magnetic field
\begin{equation}
	\rho_{x, x^{\prime}; y, y^{\prime}} \longrightarrow e^{-i\left(\chi_{x,y}-\chi_{x^{\prime},y^{\prime}}\right)}\rho_{x, x^{\prime}; y, y^{\prime}},
\end{equation}
\\
which in particular implies that the absolute value of $\rho$ is preserved.
\par
The singular lattice gauge transformation which promotes the bosonic operators to composite bosonic operators on the other hand had the form
\begin{equation}
	\hat{a}^{(\dagger)}_{x,y} \longrightarrow e^{(-)i\hat{\Phi}_{x,y}^{(\dagger)}}\hat{a}^{(\dagger)}_{x,y}.
\end{equation}  
Since $\hat{\Phi}_{x,y}$ directly depends on the site-local density operator $\hat{n}_{m,n}$ where $(m,n)\neq (x,y)$ we cannot treat it as a pure phase which we can pull out of the ensemble average 
\par
However, as $e^{(-)i\chi_{x,y}}$ is just a complex number it commutes with any $\hat{\Phi}_{x,y}$ and it follows that the composite bosonic correlation function
\begin{equation}
	\tilde{\rho}_{x,x^{\prime};y,y^{\prime}} = \left\langle^{-i\hat{\Phi}_{x,y}^{\dagger}}\hat{a}^{\dagger}_{x,y} \hat{a}^{\nodagger}_{x^{\prime},y^{\prime}} e^{i\hat{\Phi}_{x^{\prime},y^{\prime}}} \right\rangle,
\end{equation}
also simply acquires a trivial phase under gauge transformations on the physical vector potential
\begin{equation}
	\tilde{\rho}_{x,x^{\prime};y,y^{\prime}} \longrightarrow \mathrm{e}^{-i\left(\chi_{x,y}-\chi_{x^{\prime},y^{\prime}}\right)}\tilde{\rho}_{x,x^{\prime};y,y^{\prime}}.
\end{equation}
Since the singular gauge transformations is only unique up to a translation on the connected component covered by the exponential map of the group $\mathrm{U(1)}$, we can choose the extension of the singular gauge transformation to be
\begin{equation}
	\hat{\Phi}_{x,y}-\chi_{x,y}
\end{equation}
from which follows
\begin{equation}
	\tilde{\rho} \xrightarrow{\chi} \tilde{\rho}.
\end{equation}
That means, not only the norm is preserved but generally the emergence of HODLRO.
\section{Error Estimation and Convergence of the Two-Site Sampling} \label{app:A}
In this section we study the error and the convergence of the novel two-site sampling scheme. 
In principle, as the deviation from the MPS expectation value we are sampling (see Eq. \eqref{SamplingExp}) should be purely statistical, we expect a convergence of $1/\sqrt{N}$ with $N$ being the number of snapshots taken.
Following \cite{Ferris2012}, it can be readily shown that the variance $\sigma_{A}$ of the complete sampling algorithm is indeed the variance $\sigma_{\hat{A}}$ of the expectation value of the operator $\hat{A}$ we are sampling.
Complete here refers to the fact that we probe all lattice sites of the system.  
We will elaborate how this translates to the mixed one-site/ two-site scheme.
\\
Let  $\hat{A}$ be an operator on the tensor product Hilbert space $\mathcal{H}$ which consists of a sum of Hermitian site-local operators $\hat{a}_k\equiv\mathbb{1}_1\otimes \cdot \cdot \cdot \otimes \mathbb{1}_{k-1}\otimes \hat{a}_k\otimes \mathbb{1}_{k+1}\otimes \cdot \cdot \cdot \otimes \mathbb{1}_L$ on all but two sites. 
On the two remaining (adjacent) sites it consists of a Hermitian two-site tensor product operator $\hat{q}_{j,j+1}\equiv \mathbb{1}_1\otimes \cdot \cdot \cdot \otimes \mathbb{1}\otimes \hat{q}_j\otimes \hat{q}_{j+1}\otimes \mathbb{1}_{j+2}\otimes \cdot \cdot \cdot \otimes \mathbb{1}_L$ with $[\hat{a}_k,\hat{q}_{j,j+1}]=0$.
That is, we can find a basis $\{\ket{\sigma_i}\}$ in which both all site-local operators and the two-site operator is diagonal. 
Furthermore, we can express the expectation value of $\hat{A}$ regarding a state $\ket{\psi}\in \mathcal{H}$ as
\begin{equation}
	\begin{aligned}
		\bra{\psi}\hat{A}\ket{\psi} &= \sum_{\vec{\sigma}\in \Sigma} \braket{\psi|\vec{\sigma}}\bra{\vec{\sigma}}\hat{A}\ket{\psi} \\
		&= \sum_{\vec{\sigma}\in \Sigma}P(\vec{\sigma})E(\vec{\sigma}),
	\end{aligned}
\end{equation}
\begin{figure}[t]
	\centering
	\includegraphics{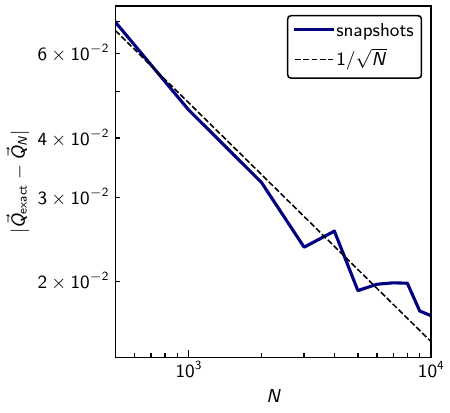}
	\caption{
		Convergence of the hybrid sampling scheme. Deviation from the quasi-exact expectation value of the snapshot averages versus number of samples $N$. $\vec{Q}_{exact}$ denotes here the vector containing all expectation values for the one-particle correlation function for a given set of lattice points. $\vec{Q}_N$ contains the corresponding values from sampling.
	}
	\label{Fig:Convergence}
\end{figure}
where we introduced both the probability $P(\vec{\sigma})$ to measure the $\ket{\psi}$ in the product state $\ket{\vec{\sigma}}=\ket{\sigma_1}\otimes \cdot \cdot \cdot \otimes \ket{\sigma_{j-1}}\otimes \left(\ket{\sigma_j}\otimes \ket{\sigma_{j+1}}\right)\otimes \ket{\sigma_{j+2}} \otimes \cdot \cdot \cdot \otimes \ket{\sigma_L}$ given by
\begin{equation}
	P(\vec{\sigma})=\braket{\psi|\vec{\sigma}}\braket{\vec{\sigma}|\psi},
\end{equation}
\begin{figure*}[ht]
	\centering
	\includegraphics{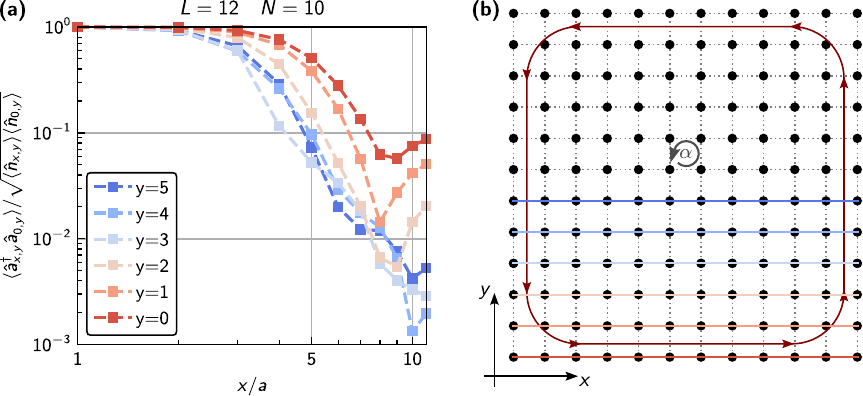}
	\caption{
		One-particle correlation function along the x-direction. Clear signatures of a slower decay close to the edge, signaling the presence of an edge current. 
	}
	\label{Fig:Edge}
\end{figure*}
and the corresponding estimator $E(\vec{\sigma})$
\begin{equation}
	E(\vec{\sigma})= \frac{\braket{\vec{\sigma}|\hat{A}|\psi}}{\braket{\vec{\sigma}|\psi}}.
\end{equation}
The mixed one-site/ two-site sampling is based on drawing $N$ independent samples from the random variable $\left(P(\vec{\sigma}),E(\vec{\sigma})\right)$.
The mean of this random variable is then by definition
\begin{equation}
	\bar{A}= \sum_{\vec{\sigma} \in \Sigma} P(\vec{\sigma})E(\vec{\sigma})=\braket{\psi|\hat{A}|\psi}.
\end{equation}
Consequently, we can also construct the variance of the variable as
\begin{equation}
	\begin{aligned}
		\sigma_A^2 &\equiv \sum_{\vec{\sigma}\in \Sigma}P(\vec{\sigma})\vert E(\vec{\sigma})-\bar{A} \vert^2 \\ &=
		\sum_{\vec{\sigma}\in \Sigma}P(\vec{\sigma})\vert E(\vec{\sigma})\vert^2 - \vert \bar{A} \vert^2=\sigma_{\hat{A}}^2
	\end{aligned}
\end{equation}
This means that the error of the approximation of the expectation value of $\hat{A}$ via drawing $N$ independent samples with standard deviation $\sigma_{A}^2$ is simply
\begin{equation}
	\begin{aligned}
		\Delta_A(N) \simeq \frac{\sigma_{\hat{A}}}{\sqrt{N}}.
	\end{aligned}
\end{equation}
 justifying the choice of the error used throughout the paper.
 \\
Additionally, we probe the convergence by comparing the MPS expectation value of the two-point correlation function for the square lattice Hofstadter-Bose-Hubbard model of size $L=12$ where we fix the reference site on the edge, i.e. $x^{\prime}=0$ at a constant $y=5$ value, i.e. we are computing $\rho_{x^{\prime}=0,y=5}(x)$ for $x\in[0,...,L-1]$. 
 We can store the expectation values in a vector $\vec{Q}_{exact}=\left(\rho_{x^{\prime}=0,y=5}(0),...,\rho_{x^{\prime}=0,y=5}(L-1)  \right)$. Analogously, we sample each of the two-point correlation functions with a certain amount of snapshots $N$ which defines $\vec{Q}_N=\left(\rho^{N}_{x^{\prime}=0,y=5}(0),..., \rho^{N}_{x^{\prime}=0,y=5}(L-1) \right)$. 
The distance between this points should then scale like $1/\sqrt{N}$, i.e. 
\begin{equation}
	\begin{aligned}
		|\vec{Q}_{exact}-\vec{Q}_N| \simeq \frac{1}{\sqrt{N}}.
	\end{aligned}
\end{equation}
In Fig. \ref{Fig:Convergence} we study this distance with increasing $N$ and find the expected convergence behavior.

\section{Edge Mode} \label{app:B}

FQH systems with open boundary conditions famously host chiral edge modes.
 In particular, the case of $\nu=1/2$ studied here is supposed to exhibit a chiral Luttinger liquid edge mode with a phase field operator that is expected to have considerable overlap with the creation and annihilation operators on the lattice. 
 \par
The existence of a chiral edge mode can be probed by considering two-point correlation functions along the edge of the system.
 While the two-point correlations are expected to decay exponentially deep in the bulk of the system, the presence of an edge mode weakens the fall-off moving closer to the edge. 
 Measuring correlations explicitly along the edge one should find an algebraic decay signaling the presence of the gapless edge mode.
Additionally,  in systems with finite size (which we are considering in our studies) one would also expect the \textit{revival} of the correlation function once we approach the opposite edge of the system~\cite{He2017}. 
 \par
 In Fig. \ref{Fig:Edge} we study the behavior of the two-point correlation functions along the $x$-direction for varying values of $y$ while keeping the reference site fixed on the edge, i.e. $x^{\prime}=0$. 
 We find that for values of $y\lesssim2$ close to the edge the decay is indeed suppressed resembling signatures of a power-law decay.
 For values of y $\simeq L/2$ deep in the bulk the correlations are vanishing exponentially. 
 Furthermore, the revival of the correlation function can be observed for all values of $y$.
 Since the edge mode has a shorter distance to cover between $x=0$ and $x=L$ we find more pronounced revivals for smaller $y$. 
Hence, we find conclusive evidence for the coexistence of a gapless edge mode living on the boundary of the system and a gapped bulk.
\begin{figure}[hb]
	\centering
	\includegraphics{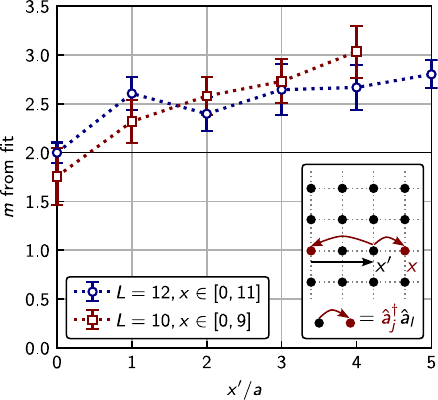}
	\caption{
		$m$ as obtained from fits of $\tilde{\rho}^{\rm LN}_{x,x^{\prime};5}$ to the data from $N_{\rm snaps}=10^4$ snapshots for varying position of the reference site $x^{\prime}$.
		The black line indicates the analytical prediction $m=2$ for the $\nu=\nicefrac{1}{2}$-Laughlin state.
	}
	\label{Fig:DependenceReferenceSite}
\end{figure}
\section{Dependence on the Reference Site} \label{app:C}
For completeness and to deepen the understanding of the edge effects in the system, we systematically change the position of the reference site for systems of size $L=10$ and $L=12$. 
That is, we compute the composite bosonic two-point correlation functions $\tilde{\rho}_{x, x^{\prime}; y}$ with $y\in \mathrm{bulk}$ for a varying $x^{\prime}\in[0,1,...,L/2]$ alongside the $x$-direction.
The immediate consequence is that the maximum distance we can reach in each of the correlations is limited by $L-x^{\prime}$.
We observe $\forall x^{\prime}$ the emergence of a power-lay decay in the  composite bosonic correlations signaling HODLRO.
\par
Quantitatively, we can study the fit through the data points giving us insight about the power-law exponent $\nicefrac{2}{m}$.
The results are shown in Fig.\ref{Fig:DependenceReferenceSite}.
For both system sizes the $m$ value obtained increases with increasing $x^{\prime}$.
We can only reproduce the correct $m$-value of continuum considerations within  the errorbars for $x^{\prime}=0$.
As discussed in the results in Section \ref{sec:Numerical Analysis}, points $x$ close to $x^{\prime}$ and points $x$ close to the edge might be effected by length scale, finite size and edge effects and are not considered for the fit.
Hence, the bigger $x^{\prime}$ the fewer data points can be taken into account and the fit quality becomes less reliable.
\section{Sampling Non-Hermitian Operators} \label{app:E}

In the context of HODLRO we are interested in correlation functions of the type 
\begin{equation}
	\left\langle \hat{a}_{\boldsymbol{j}}^{\dagger} \hat{a}_{\boldsymbol{l}}^{\nodagger} \prod_{\boldsymbol{k}\neq \boldsymbol{j},\boldsymbol{l}} \hat{n}_{\boldsymbol{k}} \right\rangle,
\end{equation}
that is the two-site operator of interest is the bosonic one-particle correlation $\hat{a}_{\boldsymbol{j}}^{\dagger}\hat{a}_{\boldsymbol{l}}^{\nodagger}$.
Clearly, this operator is non-Hermitian, i.e. $(\hat{a}^{\dagger}_{\boldsymbol{j}}\hat{a}_{\boldsymbol{l}}^{\nodagger})^{\dagger}=\hat{a}^{\dagger}_{\boldsymbol{l}}\hat{a}_{\boldsymbol{j}}^{\nodagger} \neq \hat{a}^{\dagger}_{\boldsymbol{j}}\hat{a}_{\boldsymbol{l}}^{\nodagger}$, so that the spectral decomposition used above does not exist a priori. 	
The workaround is to decompose the correlator in its Hermitian real- and imaginary-part
\begin{equation}
	\begin{aligned}
		\text{Re}(\Hat{a}_{\boldsymbol{j}}^{\dagger}\Hat{a}_{\boldsymbol{l}}^{\nodagger})&=\frac{\Hat{a}_{\boldsymbol{j}}^{\dagger}\Hat{a}_{\boldsymbol{l}}^{\nodagger}+(\Hat{a}_{\boldsymbol{j}}^{\dagger}\Hat{a}_{\boldsymbol{l}}^{\nodagger})^{\dagger}}{2},\\
		\text{Im}(\Hat{a}_{\boldsymbol{j}}^{\dagger}\Hat{a}_{\boldsymbol{l}}^{\nodagger})&=i\frac{\Hat{a}_{\boldsymbol{j}}^{\dagger}\Hat{a}_{\boldsymbol{l}}^{\nodagger}-(\Hat{a}_{\boldsymbol{j}}^{\dagger}\Hat{a}_{\boldsymbol{l}}^{\nodagger})^{\dagger}}{2}.
	\end{aligned}
\end{equation}
But there is another caveat. 
The real- and the imaginary part are not commuting simply because the creation and annihilation operators on the same site-local Hilbert space fulfill
\begin{equation}
	\begin{aligned}
		[\Hat{a}_{\boldsymbol{j}}^{\nodagger},\Hat{a}_{\boldsymbol{j}}^{\dagger}] \neq 0,
	\end{aligned}
\end{equation}
from where it immediately follows that
\begin{equation}
	\begin{aligned}
		[\text{Re}(\Hat{a}_{\boldsymbol{j}}^{\dagger}\Hat{a}_{\boldsymbol{l}}^{\nodagger}),\text{Im}(\Hat{a}_{\boldsymbol{j}}^{\dagger}\Hat{a}_{\boldsymbol{l}}^{\nodagger})] \neq 0.
	\end{aligned}
\end{equation}
This implies that we cannot simultaneously diagonalize them, i.e. we cannot sample both in the same sweep. 
That forces us to perform two (quasi)-independent sweeps over the lattice: one where we sample the real-part, and one where we sample the imaginary-part.
\par
In order to obtain physically compatible measurements, we need to assure that for both the real- and imaginary part the Fock-state snapshots agree on the other sites. This is done in the following way:
\begin{enumerate}
	\item Draw the same random number at each sampling step for both sweeps.
	\item Sample first all particle position.
\end{enumerate}
The first argument ensures that we draw the same eigenvalue for each site-local tensor, if the probability distribution coincides.
The equality of the probability distribution for the Fock-state samples is guaranteed by the second condition. 
We can achieve this by sweeping first from left-to-right up until site $\boldsymbol{j}-1$. 
Then we left-normalize $\ket{\psi_{\boldsymbol{j}-1}}$, such that the canonical center is the $L$th site. 
From thereon, we sweep from right-to-left until we reach site $\boldsymbol{j}+2$. 
Ultimately, we move the canonical center to $(\boldsymbol{j},\boldsymbol{j}+1)$ and sample $\Hat{a}_{\boldsymbol{j}}^{\dagger}\hat{a}_{\boldsymbol{j}+1}^{\nodagger}$.
The conditioned probability distribution is then given by
\begin{equation}
	\begin{aligned}
		P(p_{\boldsymbol{j},\boldsymbol{j}+1}/\tilde{p}_{\boldsymbol{j},\boldsymbol{j}+1}|n_{\boldsymbol{j}+2},...,n_{\boldsymbol{L}},n_{\boldsymbol{j}-1},...,n_{\boldsymbol{1}}).
	\end{aligned}
\end{equation}
Here we used $p_{\boldsymbol{j},\boldsymbol{j}+1}/ \tilde{p}_{\boldsymbol{j},\boldsymbol{j}+1}$ and $n_{\boldsymbol{k}}$ to denote the drawn eigenvalues of the real- respectively the imaginary-part and the local Fock-states.
Now suppose we perform a single sampling step, as described above, yielding the snapshot
\begin{equation}
	\begin{aligned}
		\text{Real part:} \ &(n_{\boldsymbol{1}},...,n_{\boldsymbol{j}-1},p_{\boldsymbol{j},\boldsymbol{j}+1},n_{\boldsymbol{j}+2},...,n_{\boldsymbol{L}}) \\
		\text{Imag. part:} \ &(n_{\boldsymbol{1}},...,n_{\boldsymbol{j}-1},\tilde{p}_{\boldsymbol{j},\boldsymbol{j}+1},n_{\boldsymbol{j}+2},...,n_{\boldsymbol{L}})
	\end{aligned}
\end{equation}
The local densities coincide. 

\section{Extracting HODLRO Snapshots} \label{app:F}
In this paper, we applied the described algorithm to numerically approximate the expectation value of the composite boson correlator, i.e.
\begin{equation}\label{HODLRO obs}
	\begin{aligned}
		\left\langle \prod_{\boldsymbol{k} \neq \boldsymbol{j},\boldsymbol{j}+1} \left( \frac{z_{\boldsymbol{j}} - z_{\boldsymbol{k}} }{|z_{\boldsymbol{j}} - z_{\boldsymbol{k}}|} \right)^{-m \hat{n}_{\boldsymbol{k}}} \left( \frac{z_{\boldsymbol{j}+1}- z_{\boldsymbol{k}} }{|z_{\boldsymbol{j}+1} - z_{\boldsymbol{k}}|} \right)^{m \hat{n}_{\boldsymbol{k}}} \hat{a}^{\dagger}_{\boldsymbol{j}} \hat{a}^{\nodagger}_{\boldsymbol{j}+1} \right\rangle.
	\end{aligned}
\end{equation}
We will briefly explore how to evaluate Eq.~\eqref{HODLRO obs} by using the one-site/ two-site sampling scheme.
For a lattice system with N particles the set $\{n_{\boldsymbol{k}}\}$ with $
k\in[\boldsymbol{1},\boldsymbol{2},...,\boldsymbol{j}-1,\boldsymbol{j}+2,...,\boldsymbol{L}]$ fulfills the condition
\begin{equation} \label{Rho:zero}
	\begin{aligned}
		&\sum_{n_{\boldsymbol{k}}} n_{\boldsymbol{k}} = N \iff \tilde{p}_{\boldsymbol{j},\boldsymbol{j}+1}=p_{\boldsymbol{j},\boldsymbol{j}+1}=0,\\
		&\sum_{n_{\boldsymbol{k}}} n_{\boldsymbol{k}} = N-1 \iff \tilde{p}_{\boldsymbol{j},\boldsymbol{j}+1} \neq0 \neq p_{\boldsymbol{j},\boldsymbol{j}+1}.
	\end{aligned}
\end{equation}
In the hardcore bosonic case the local occupations are constrained to $n_{\boldsymbol{k}}\in\{0,1\}$ and we can define:
\begin{equation}
	\begin{aligned}
		n_{\boldsymbol{k}}=1 \Rightarrow n_{\boldsymbol{k}}\equiv \tilde{n}_l \ \ \ , \ \ \ l\in\{1,2,...,N-1\}.
	\end{aligned}
\end{equation}
The index $l$ is just running up until $N-1$, because we can neglect the case where we find $N$ particles, due to Eq. \eqref{Rho:zero}. 
Each non-zero Fock basis snapshot $n_{\boldsymbol{k}}$ is then contributing as a non trivial phase and the single one-particle composite boson correlator snapshot reads
\begin{equation}
	\begin{aligned} 
		\prod_{l=1}^{N-1}\left( \frac{z_{\boldsymbol{j}} - z_l }{|z_{\boldsymbol{j}} - z_l|} \right)^{-m} & \left( \frac{z_{\boldsymbol{j}+1}- z_l }{|z_{\boldsymbol{j}+1} - z_l|} \right)^{m} \\ \times (p_{\boldsymbol{j},\boldsymbol{j}+1}+&i\tilde{p}_{\boldsymbol{j},\boldsymbol{j}+1}).
	\end{aligned}
\end{equation}
\\

\end{document}